\documentclass[prb]{revtex4}% Physical Review B
\usepackage{graphicx}% Include figure files
\usepackage{dcolumn}% Align table columns on decimal point
\usepackage{bm}% bold math
\usepackage{subfigure}
\begin{document}
%7 \preprint{APS/123-QED}
\title{Spatio-temporal Bounded Noises, and transitions induced by them in solutions of real Ginzburg-Landau model}% Force line breaks with \\
\author{Sebastiano de Franciscis}
\affiliation{
European Institute of Oncology, Department of Experimental Oncology,Via Ripamonti 435, I20141 Milano (Italy)}%
\author{Alberto d'Onofrio}
\email{alberto.donofrio@ifom-ieo-campus.it}
\affiliation{(Corresponding author)\\European Institute of Oncology, Department of Experimental Oncology,Via Ripamonti 435, I20141 Milano (Italy)}%
\date{\today}% It is always \today, today,
% but any date may be explicitly specified
\begin{abstract} In this work, we introduce two spatio-temporal colored bounded noises, based on the zero-dimensional Cai-Lin and Tsallis-Borland noises.\\ We then study and characterize the dependence of the defined bounded noises on both a temporal correlation parameter $\tau$ and on a spatial coupling parameter $\lambda$. The boundedness of these noises has some consequences on their equilibrium distributions. Indeed in some cases varying $\lambda$ may induce a transition of the distribution of the noise from bimodality to unimodality. \\
With the aim to study the role played by bounded noises on nonlinear dynamical systems, we investigate the behavior of the real Ginzburg-Landau time-varying model additively perturbed by such noises. The observed phase transitions phenomenology is quite different from the one observed when the perturbations are unbounded.\\
In particular, we observed an inverse "order-to-disorder" transition, and a re-entrant transition, with dependence on the specific type of bounded noise.
\end{abstract}
%Use showkeys class option if keyword
\maketitle
\section{Introduction}
Since the hallmark seminal works on Brownian motion by Einstein and Langevin, Gaussian noises (GNs) have been one of the main concepts used in non-equilibrium statistical physics, and one of the main tools of its applications, from engineering to biology.
The later, and quite dichotomic, mathematical works by Ito and Stratonovich laid a firm theoretical basis for the mathematical theory of stochastic differential equations, as well as a long-lasting controversy on which of the two approaches is the best suited in order to better describe stochastic fluctuations in the real world. 
Other hallmarks in stochastic physics have been introduced in the seventies with the birth, in the framework of the Ilya Prigogine school, of the theory of noise-induced transitions by Horsthemke and Lefever \cite{hl} and in the early eighties, in the context of the Rome school, with the introduction of the concept of stochastic resonance \cite{lucafrancesco}. The study of stochastic perturbations affecting excitable systems revealed the existence of another stochastic counter-intuitive effect: the coherence-resonance \cite{lucafrancesco}. Finally, in last 20 years nonlinear analysts developed a a rigorous theory of stochastic bifurcations – both phenomenological and dynamics \cite{arnold}.\\
Both stochastic resonance and coherence-resonance are example of noise-induced emergence of ordered structures in zero-dimensional systems \cite{gammaitoni, lucafrancesco}. In other words, noise in a nonlinear system is able to create order. This concept stimulated a large amount of research on constructive stochastic effects in spatially extended systems. A first important result of these investigations has been the emergence of noise-induced phase transitions "disorder-to-order" in spatially extended systems. In those cases a key bifurcation parameter is the strength of the spatial coupling (e.g. in the case of Laplace coupling, the diffusion coefficient). It is remarkable that this phenomenology was also observed in the case of stochastic perturbations of deterministic physical systems unable to form patterns, i.e. when moving from a spatially homogeneous to a spatially coupled deterministic system there is no emergence of Turing patterns. The addition of noise to such systems may cause the onset of "disorder-to-order" phase transitions. An example of such a model is the Ginzburg-Landau (GL) real equation \cite{onuki, sancho, Ibanhes}.\\
The majority of works in the field of noise-induced phase transitions were based on white noises, in which is absent any spatio-temporal correlation or structure. This kind of fluctuations is appropriate when modeling internal "hidden" degrees of freedom, of microscopic dynamical nature. On the contrary, fluctuations originated externally to the system in study may exhibit both temporal and spatial structure \cite{GObook,sanchoPhysD}. Zero-dimensional systems perturbed by colored Ornstein-Uhlenbeck noise (OU) exhibit correlation-dependent properties that are missing in case of null autocorrelation. Among them: the emergence of stochastic resonance also for linear systems, and re-entrance phenomena, i.e. transitions from monostability to bistability and back to monostability \cite{wiolindenberg,hanggi,hanggi2}. Spatially extended systems exhibit even more striking effects when they are perturbed by spatially white but temporally colored noises, due to a complex interplay between noise intensity, spatial coupling (of the perturbed system) and autocorrelation time \cite{wiolindenberg}.\\
Garc\'{i}a-Ojalvo, Sancho and Ram\'{i}rez-Piscina introduced in \cite{GO92} the spatial version of the Ornstein-Uhlenbeck noise, characterized by both a temporal scale $\tau$ and by a spatial scale $\lambda$, and studied analytically the spatio-temporal correlation of this kind of noise in \cite{lam}. Later they showed that the interplay of this noise with suitable nonlinear systems can induce transitions often unobserved when the applied noise is white. In particular, when analyzing the additive spatio-temporally colored perturbation of the Ginzburg-Landau field, they showed the existence of a nonequilibrium phase transition controlled by both the correlation time and the correlation length \cite{GO94,GObook}. Similarly to the deterministic bifurcations ruled by two parameters, the transitions induced by the Garc\'{i}a-Ojalvo, Sancho and Ram\'{i}rez-Piscina spatio-temporal noise (GSR) show a more complex picture than the transitions induced by a spatio-temporal noise that is spatially uncorrelated \cite{GO94}. In the case of a single spatial dimension, a different extension of the Ornstein-Uhlenbeck noise was proposed, and analytically studied in \cite{traulsen}. \\
The above-summarized body of research is essentially based on the use of GNs, whose background is the Central Limit Theorem, and which is the best approximation of reality in many cases. 
However, an increasing number of experimental data motivated theoretical studies stressing that many real-life stochastic processes does not follow white or colored Gaussian laws, but other densities (such as “fat-tail” power-laws). More recently, a vast body of research focused on another important class of non-Gaussian stochastic processes: the bounded noises. 
The studies on bounded noises, apart some sporadic exception, were mainly confined to the Dichotomous Markov Noise (DMN) and to its applications \cite{lucafrancesco}. In the last twenty years, together with a renewal of theoretical interest for DMN \cite{lucafrancesco}, other classes of bounded noises were defined and intensively studied in statistical physics and in engineering, and - to a lesser degree - in mathematics and quantitative biology.\\
The rise of scientific interest on bounded noises is motivated by the fact that in many applications both GNs and “fat-tailed” non-Gaussian stochastic processes are an inadequate mathematical model of the physical world because of their infinite domain. This should preclude their use to model stochastic fluctuations affecting parameters of dynamical systems, which must be bounded by physical constraints. Moreover, in many relevant cases, especially in biology, some parameters must also be strictly positive. As a consequence, not taking into account the bounded nature of stochastic fluctuations may lead to unrealistic inferences. For instance, when the onset of noise-induced transitions depends on exceeding a threshold by the variance of a GN, this often means making negative or excessively large a parameter. To give an example taken from medicine, a GN-based modeling of the unavoidable fluctuations affecting the pharmacokinetics of an antitumor drug delivered by means of continuous infusion leads to a paradox. Indeed, the probability that the drug increases the number of tumor cells may become nonzero, which is absurd \cite{dongan}. Thus in order to avoid these problems, the stochastic models should in these cases be built on bounded noises.\\
The deepening and development of theoretical studies on bounded noises led to the attention of a vast readership on new phenomena, such as the dependence of the transitions on the specific model of noise that has been adopted\cite{pre,dongan}. This means that, in absence of experimental data on the density and spectrum of the stochastic fluctuations for the problem in study, a scientific work should compare multiple kinds of possible stochastic perturbations. Moreover, currently the bounded noise approach also implies that the possibility of obtaining analytical results is remarkably reduced. For example, the study of stochastic bounded perturbations acting on a simple scalar deterministic model leads to at least
two stochastic differential equations. Indeed, one or more additional equations must be devoted to the modeling of the bounded stochastic processes.\\
Finally, we remind that in order to generate a temporal bounded noise, two basic recipes have been adopted so far. The first consists in applying a bounded function to a random walk \cite{bobryk}, whereas the second one consists in generating the noise by means of an appropriate stochastic differential equation\cite{wioII,CaiLin}.\\
Our aim here is twofold. First, we want to define two simple families of spatio-temporally bounded noises, which extend two kinds of temporal noises frequently employed in literature, the Tsallis-Borland noise \cite{wioII}, and the Cai-Lin noise \cite{CaiLin}.\\ 
Second, we want to illustrate the possible effects of external bounded stochastic forces (i.e. of additive bounded noises) on nonlinear spatio-temporal models. As a case study we have chosen the scalar Ginzburg-Landau equation, one of the best-studied amplitude equation representing "universal" nonlinear mechanisms.\\ Phase transitions induced in GL model by additive and multiplicative unbounded noises were extensively studied for this model \cite{GObook,GO92pla,GO92,gpsv,various,ms,jstat,ss,lucafrancesco}. GL equation driven by temporally varying Markov dichotomous noise have been considered in \cite{ouch}. Moreover, recently it has been investigated the case of GL dynamics under external additive "quenched noise" (i.e. a random perturbation that is very slow in comparison to the unperturbed model dynamics) has been studied in \cite{jstat}, where the possibility of multistability was analytically studied.\\
Simulations with the GSR unbounded noise will be performed, in order to dissect the effects that are more closely related to the boundedness of the noises, and to their type.\\
Among the tools used for the analysis of the numerical simulations, we shall also use the Moran index \cite{moran,griffith,Atlas}, a classical tool of spatial statistics of lattice data, which has been intensively applied in geographical statistics \cite{Atlas} and in image processing \cite{Atlas}, which - however - has never been used in statistical physics, at the best of our knowledge.
\section{Background on temporal bounded noise}\label{BackCaiTsallis}
Here we provide some background material related to the generation of temporally correlated bounded noises. 
\subsection{The Tsallis-Borland bounded noise}
The first noise we take in consideration is the Tsallis-Borland bounded noise \cite{wio, wioII}, whose dynamics is described by the following equation:
\begin{equation}\label{tbw}
\xi^{\prime} (t)= -\frac{1}{\tau}\frac{\xi(t)}{1-(\xi(t)/B)^2} + \frac{\sqrt {2 D} }{\tau}\eta(t).
\end{equation}
A Tsallis q-statistics\cite{wio, wioII} is the stationary distribution of this noise:
\begin{equation}\label{Peqtsallis}
P_{TB}(\xi)= A (B^2-\xi^2)_+^{\frac{1}{1-q}} ,
\end{equation}
whose mean is null, and whose standard deviation is given by
$$ \sigma_{TB}(q) = B \frac{\Gamma \left(\frac{3}{2}+\frac{1}{1-q}\right)}{2\Gamma\left(\frac{5}{2}+\frac{1}{1-q}\right)}. $$
The parameter $q \in \left[-\infty,1\right] $ links the true autocorrelation time $\tau_c$ of $\xi(t)$ to the parameter $\tau$ as follows \cite{wioII}:
$$ \tau \approx \tau_c \frac{5-3 q}{2}, $$
while the coefficient $D$ is in turn linked to the bound $B$ and to the Tsallis parameter $q$ by the following relationship:
$$ \frac{\sqrt {2 D} }{\tau} = B \sqrt{\frac{2}{\tau_C}\frac{1-q}{5-3 q} }. $$
In the Supplementary Materials (SMs) we define a generalization of the Tsallis-Borland noise. 
\subsection{The Cai-Lin bounded noise}
In \cite{CaiLin, CaiLinII, caiwu}, the following family of bounded noises was introduced:
\begin{equation}\label{Cai}
\xi^{\prime} (t)= -\frac{1}{\tau_c}\xi(t) + g(\xi)\eta(t),
\end{equation}
with $g(|B|)=0$. The bounded nature of the noise described in \ref{Cai} easily follows from the fact that at $\xi=+B$ it is $\xi^{\prime}<0$, whereas at
$\xi= - B$ it is $\xi^{\prime}>0$. \\
Note that if $g(\xi)$ is symmetric then the process $\xi(t)$ has zero mean, and the same autocorrelation of the OU process \cite{CaiLin, CaiLinII}, i.e. $\tau_c$ denotes the real autocorrelation time of the process $\xi(t)$.\\
Observe that, as shown in \cite{CaiLin, CaiLinII}, a pre-assigned generic stationary $P(\xi)$ can be obtained by the eq. (\ref{Cai}). Indeed, by the Fokker-Planck equation associated to eq. (\ref{Cai}) it is straightforward to verify that the following function $g(\xi)$:
$$
g(\xi)=\sqrt{ -\frac{2}{\tau_cP(\xi)}\int_{-B}^{x}uP(u)du}.
$$
has to be chosen in order to recover $P(\xi)$. For example, by choosing the following symmetric function\cite{CaiLin, CaiLinII}:
\begin{equation}\label{CaiG}
g(\xi)= \sqrt{\frac{1}{\tau_c(1+\delta)}}\sqrt{B^2 -\xi^2},
\end{equation}
one gets the following stationary density:
\begin{equation}\label{Peqcai}
P_{CL}(\xi) = A \left(B^2 - \xi^2\right)_{+}^{\delta},
\end{equation}
parametrized by $\delta>-1$, whose standard deviation is given by:
$$ 
\sigma_{CL}(\delta)= B\frac{\Gamma \left(\delta +\frac{3}{2}\right)}{2 \Gamma\left(\delta+\frac{5}{2}\right)}.
$$
For $\delta>0$ the distribution is unimodal and centered in $0$, while for $-1<\delta<0$ it is bimodal, having a "horns"-like distribution with two vertical asymptotes at $\xi\rightarrow \pm B$.
\section{spatio-temporal colored noise: background and new definitions}\label{stn} 
Let us consider the well-known Ornstein-Uhlenbeck equation:
\begin{equation}\label{oue}
\xi^{\prime} (t)= -\frac{1}{\tau_c}\xi(t) + \frac{\sqrt {2 D} }{\tau_c}\eta(t),
\end{equation}
where $\eta(t)$ is a Gaussian white noise of unitary intensity:$ \langle \eta(t)\eta(t-t_1)\rangle$ $=$ $\delta(t-t_1) $,
and whose solution is a Gaussian colored stochastic process with autocorrelation:
$$ \langle \eta(t)\eta(t-t_1)\rangle = A\ exp\left(-\frac{|t-t_1|}{\tau_c}\right).$$
Equation (\ref{oue}) has been generalized in \cite{GO92pla} as follows:
\begin{equation}\label{oupde}
\partial_t \xi (x,t)= -\frac{1}{\tau_c}\xi(x,t) + \frac{\sqrt {2 D} }{\tau_c}\eta(x,t),
\end{equation}
where $\eta(x,t)$ is a noise white in space and time. As a consequence $\xi (x,t)$ is a stochastic process white in space but colored in the temporal dimension. Finally, a noise also colored in space was proposed in \cite{GO92pla}, by adding to eq. (\ref{oupde}) the most known and simple spatial coupling, the Laplace operator, thus yielding the following partial differential Langevin equation
\begin{equation}\label{gener}
\partial_t \xi (x,t)= \frac{2\lambda^2}{\tau_c}\nabla^2 \xi(x,t) -\frac{1}{\tau_c}\xi(x,t) + \frac{\sqrt {2 D} }{\tau_c}\eta(x,t),
\end{equation}
where $\lambda>0$ is the spatial correlation length \cite{GO92} of $\xi (x,t)$. As usual in non-equilibrium statistical physics, in line with \cite{GO92pla, GO92}, we shall investigate the lattice version of eq. (\ref{gener}), i.e.:
\begin{equation}\label{generlattice}
\xi_p^{\prime} (t)= \frac{\lambda^2}{2\tau_c}\nabla_L^2 \xi_p(t) -\frac{1}{\tau_c}\xi_p(t) + \frac{\sqrt {2 D} }{\tau_c}\eta_p(t),
\end{equation}
where $p = h \ x (i,j) $ is a point on a $M*M$ lattice with steps equal to $h$. The symbol $\nabla_L^2$ denotes the discrete version of the Laplace's operator:
\begin{equation}\label {lapllatt}
\nabla_L^2 \xi_p (t)= h^{-2}\sum_{i \in Ne(p)}(\phi_i-\phi_p),
\end{equation}
where $Ne(p)$ is the set of the neighbors of the lattice point $p$.\\
In order to define spatio-temporal bounded noises based on the Tsallis-Borland and on the Cai-Lin noises, we shall use an approach analogous to the one employed in \cite{GO92pla,GO92} to extend the OU process.\\
Thus the Tsallis noise can be generalized to spatially extended systems by means of the equation:
\begin{equation}\label {sptTB}
\partial_t \xi (x,t)= \frac{\lambda^2}{2\tau_c}\nabla^2 \xi(x,t) -\frac{1}{\tau}\frac{\xi(t)}{1-(\xi(t)/B)^2} + \frac{\sqrt {2 D} }{\tau}\eta(x,t),
\end{equation}
while the spatio-temporal Cai-Lin noise can be defined as follows:
\begin{equation}\label {sptCai}
\partial_t \xi (x,t)= \frac{2\lambda^2}{\tau_c}\nabla^2 \xi(x,t) -\frac{1}{\tau_c}\xi(x,t) + \sqrt{\frac{2D}{\tau_c(1+\delta)}}\sqrt{B^2 -\xi^2}\eta(x,t).
\end{equation}
We shall study the above two equations in their lattice-based discretizations.
%which on a lattice reads:
%\begin{equation}\label {sptCailattice}
%\xi_p^{\prime} (t)= \frac{2\lambda^2}{\tau}\nabla_L^2 \xi_p(t) -\frac{1}{\tau}\xi_p + \frac{\sqrt {2 D} }{\tau}g(\xi_p)\eta_p(t).
%\end{equation}
\section{Statistical features of spatio-temporal noise}\label{StuNoise}
\begin{figure}[b]
\begin{center} 
\subfigure[]
{
\label{A}
\includegraphics[width=7.5cm]{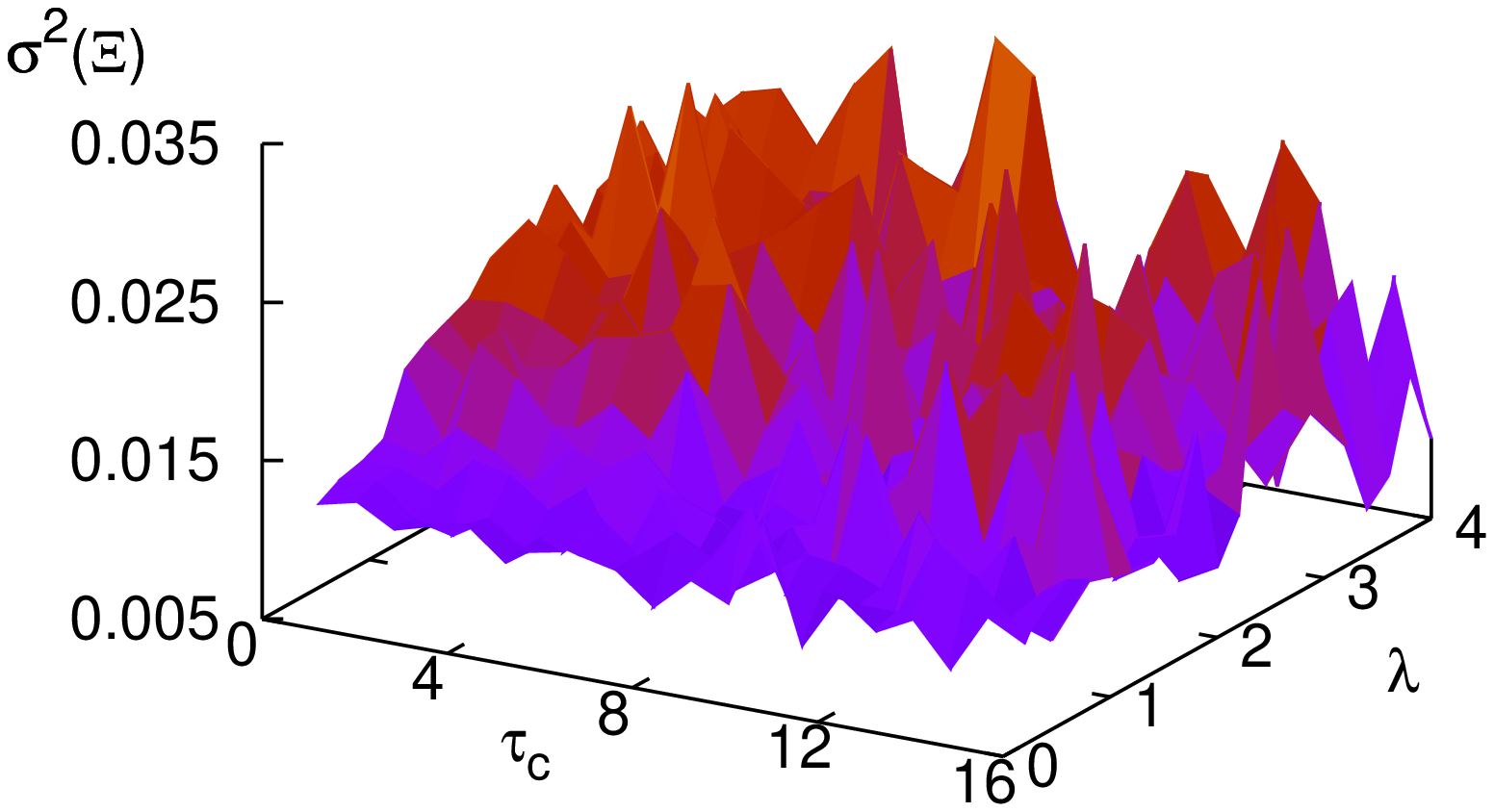}
}
\subfigure[]
{
\label{B}
\includegraphics[width=7.5cm]{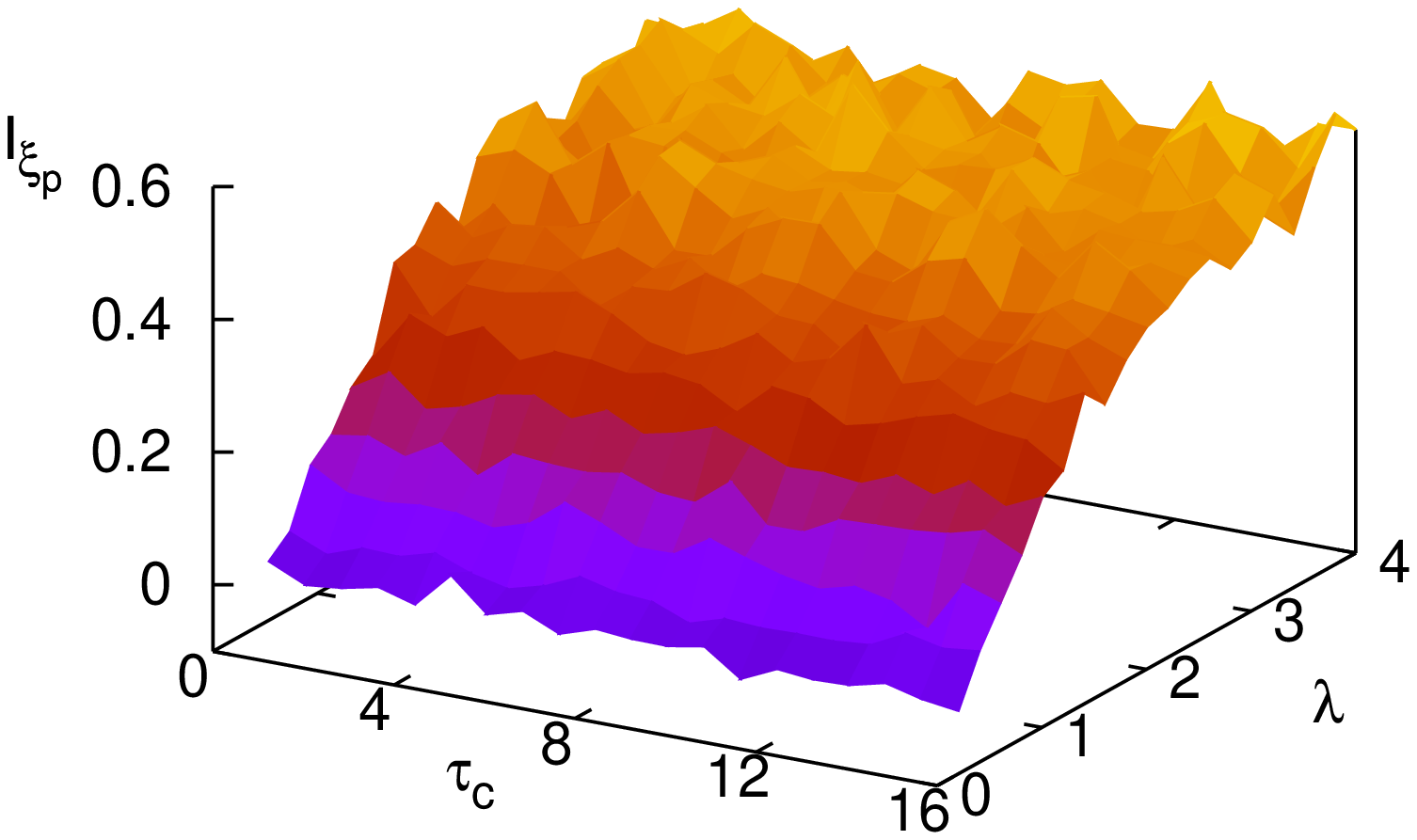}
}
%\subfigure[]
%{
%\label{C}
%\includegraphics[width=7.5cm]{GarOjaCaidm0p5sigma.eps}
%}
%\subfigure[]
%{
%\label{D}
%\includegraphics[width=7.5cm]{GarOjaCaidm0p5Moran.eps}
%}
\end{center}
\caption{Tsallis-Borland bounded noise : characterization of spatio-temporal fluctuations. Panel (a) plot of $\sigma\left(\Xi\right)$; panel (b): Moran index. Parameters: $40\times40$ lattice, $B=1$ and $q=-1$.}
\label{fig_Bound}
\end{figure}

\begin{figure}[htb!]
\begin{center}
\subfigure[]
{
\label{A}
\includegraphics[width=7.5cm]{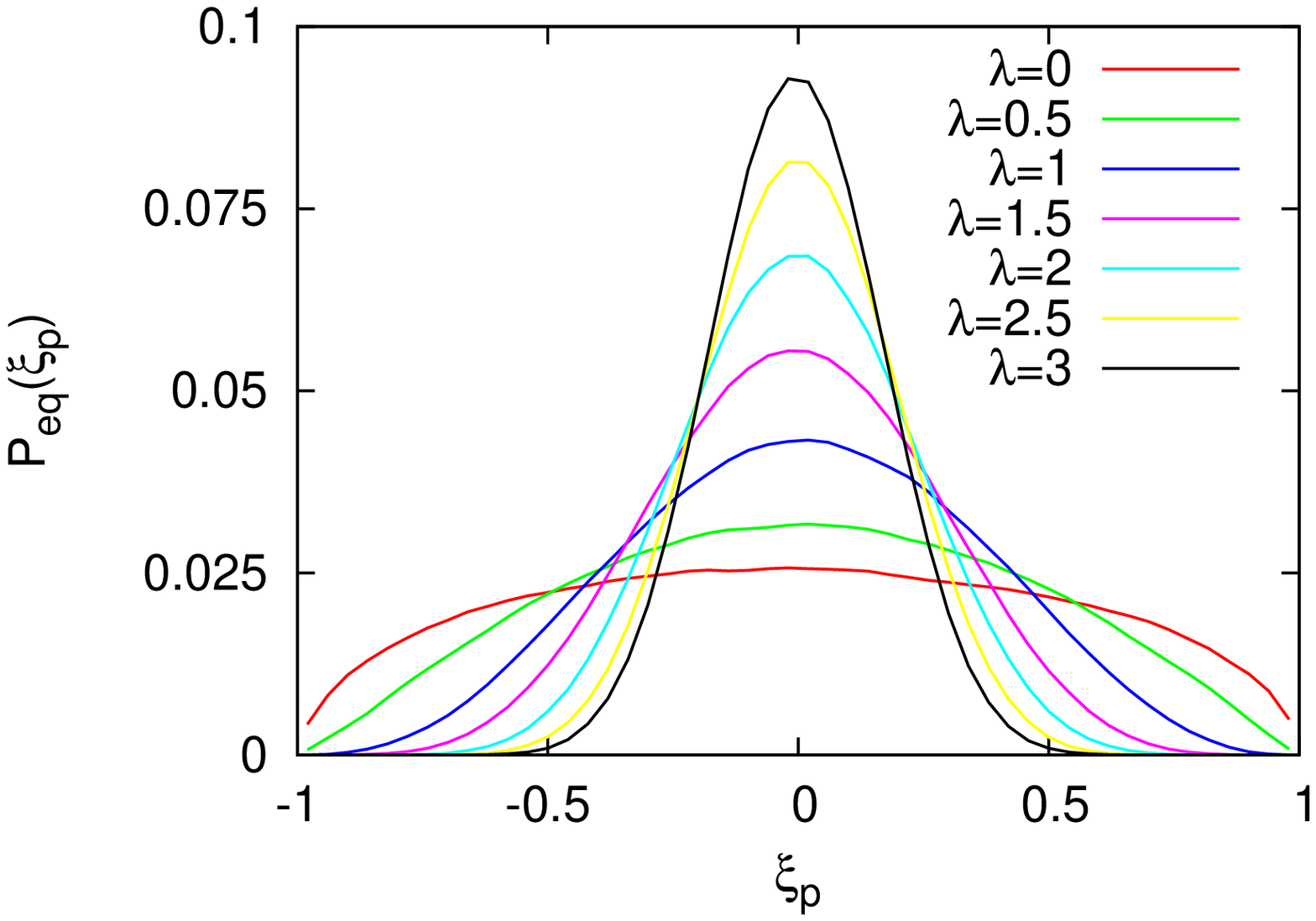}
}
\subfigure[]
{
\label{B}
\includegraphics[width=7.5cm]{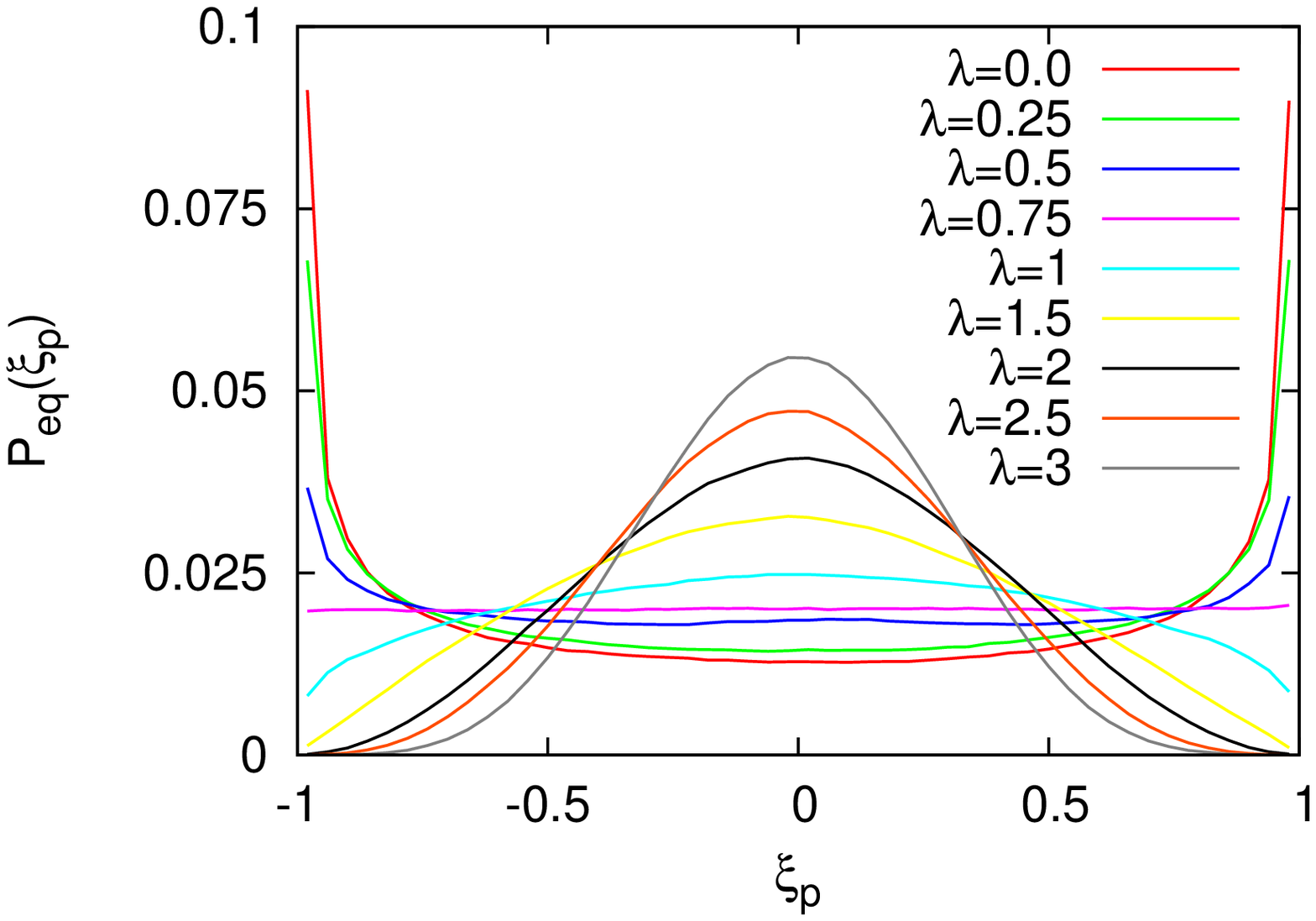}
}
%\subfigure[]
%{
%\label{C}
%\includegraphics[width=7.5cm]{TransTau2qm1.eps}
%}
\end{center}
\caption{Influence of characteristic correlation length $\lambda$ on Cai bounded noise equilibrium distribution $P_{eq}(\xi_p)$ of a $40\times40$ lattice system. Panel \subref{A}: parameters $B=1$, $\tau=2$ and $\delta=+0.5$. Panel \subref{B}: same noise with $\tau=1$ and $\delta=-0.5$.
% Panel \subref{C}: Tsallis noise with $B=1$, $\tau=2$ and $q=-1$.
}
\label{fig_P_eq_tauconst}
\end{figure}

We mainly characterized the global behavior of the bounded nosies defined by the lattice versions of eqs. (\ref{sptTB}) and (\ref{sptCai}) by means of the well known order parameter \cite{GObook} 
$$\Xi\equiv\frac{<|\sum_{(i, j)}\xi_{i,j}|>}{N^{2}},$$ and of its relative fluctuations $\sigma^2\left(\Xi\right)$, defined as \cite{GObook} :
$$
\sigma^2\left(\Xi\right)\equiv \frac{<|\sum_{(i, j)}\xi_{i,j}|^{2}>-<|\sum_{(i, j)}\xi_{i,j}|>^{2}}{N^{2}} .
$$
Concerning, instead, the characterization of the spatial-autocorrelation-related properties of the spatio-temporal noises treated in section \ref{stn}, and in the study of phase transitions induced by them, we employed a classical tool of spatial statistics: the Moran's index \cite{moran,griffith}. Given a field $\alpha_i$ on a lattice, let us define the following field $\chi_i$ on the same lattice:
\begin{equation}\label{psi}
\chi_i(\alpha) = \frac{1}{NNe(i)}\sum_{i \in Ne(i)}\alpha_i = +\alpha_i + h^2 \nabla^2_L \alpha_i,
\end{equation}
the Moran's statistic $I_ {\alpha}$ is defined as follows:
\begin{equation}\label{moranindex}
I_ {\alpha}= PC\left( \alpha, \chi(\alpha) \right),
\end{equation}
where $PC(u,v)$ is the Pearson correlation between two vectors $u$ and $v$. Moreover, in order to compare the degree of mutual correlation of two different lattice fields $\alpha$ and $\gamma$, some years ago we defined a bivariate Moran index \cite{Atlas}:
\begin{equation}\label{bivarmoranindex}
I_ {\alpha\gamma}= PC\left( \chi(\alpha_*), \chi(\gamma_*) \right),
\end{equation}
where the $*$ in the pedix denotes the normalization of the lattice field, e.g.: $\alpha_* $ $=$ $(\alpha - \langle \alpha \rangle)/\sqrt{Var(\alpha)}$.\\
We first investigated the dependence of the bounded spatio-temporal noises defined by eq. (\ref{sptTB})-(\ref{sptCai}) on the parameters $\tau_C$ (measuring the temporal autocorrelation) and $\lambda$ (measuring the spatial coupling). For both Tsallis-Borland and Cai-Lin noises, in our simulations we obtained that the fluctuation $\sigma^2$ is independent of both $\tau_C$ and $\lambda$ (see fig. \ref{fig_Bound}.a and 1.a of SM). On the contrary, as expected, the Moran index strongly depends on $\lambda$, and it is independent on $\tau_C$ (see fig. \ref{fig_Bound}.b and 1.b of SM). 
We also simulated the GSR unbounded noise by empirically setting (in order to get a noise strength somewhat comparable to the one employed in the bounded cases) $ B = 2 \sigma_u$, where $\sigma_u=\sqrt{2D}$ is the strength of the GSR noise, i.e. the standard deviation of the generative white noise in equation (\ref{oue}). We obtained that the fluctuation parameter $\sigma^2$ in this case is quite large and it is -differently from the bounded noise case - a decreasing function of $\tau$ (see fig. 2 of SM). Instead, the Moran index behaves as in the previous case.\\
%\FloatBarrier
The global behavior of the spatio-temporal noise, analyzed by the macroscopic observable $\Xi$, has its counterpart in the equilibrium heuristic probability density of the lattice variables $\xi_p$. First, we observed that in both types of bounded noises, the distribution of $\xi_p$ is independent of the temporal correlation parameter (see figures 3 and 4.b of SM%\ref{fig_P_eq_lambdaconst}
). On the contrary, the spatial coupling parameter $\lambda$ deeply affects the distribution of $\xi_p$, and in a noise-type dependent manner. Indeed: 
\begin{itemize}
\item for Cai noise with $\delta<0$, the increase of $\lambda$ induces a transition from a bimodal density to a unimodal density (see panel B of figure \ref{fig_P_eq_tauconst}); 
\item for both Cai noise with $\delta>0$ and for Tsallis noise, the increase of the spatial coupling induces a decrease of the variance of the noise (see fig. \ref{fig_P_eq_tauconst} and figure 4.a of SM).
\end{itemize}
\section{Bounded-noise-induced phase transitions in time-dependent Ginzburg-Landau model}
\begin{figure}[b]
\begin{center}
\subfigure[]
{
\label{A}
\includegraphics[width=7.5cm]{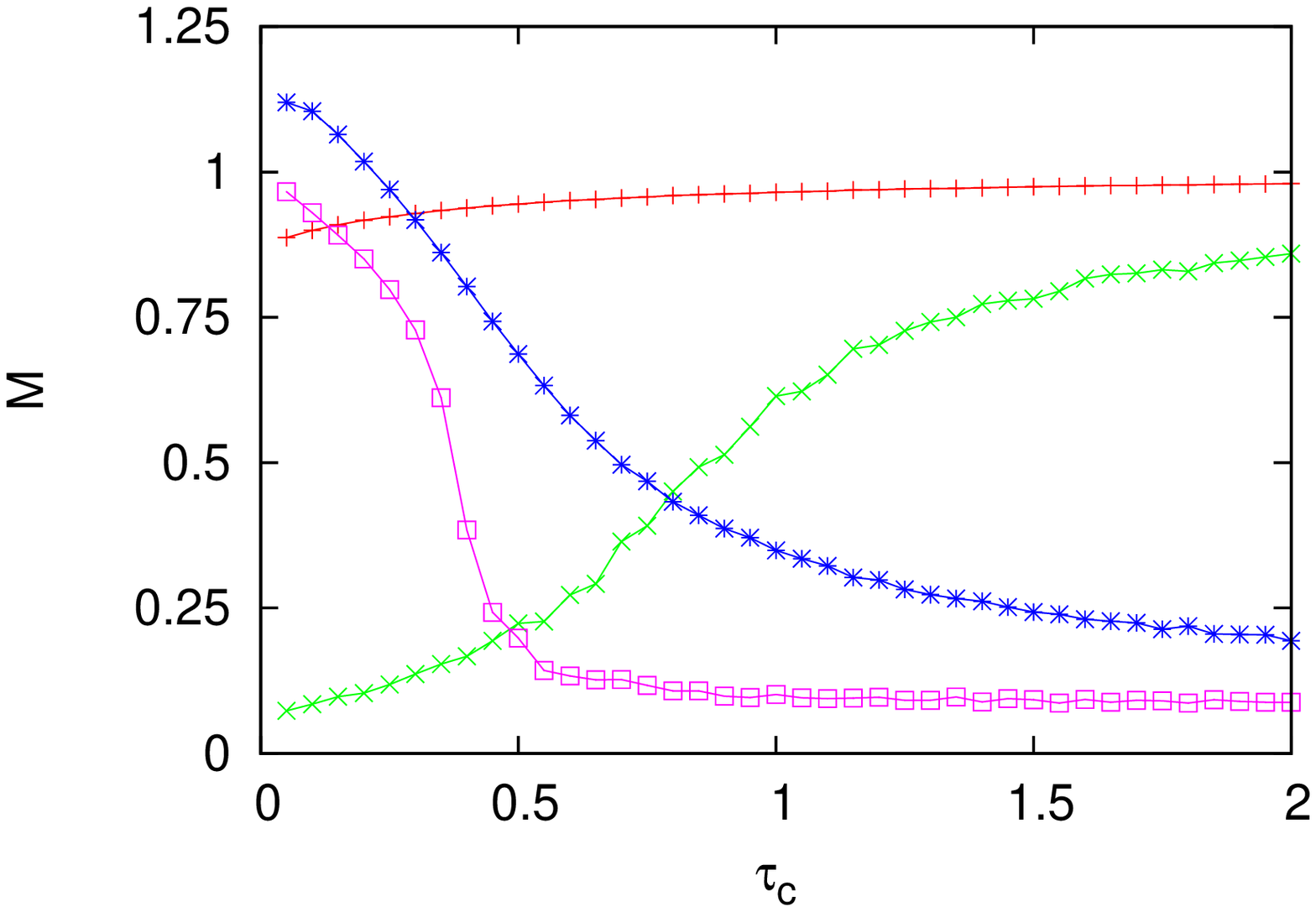}
}
\subfigure[]
{
\label{B}
\includegraphics[width=7.5cm]{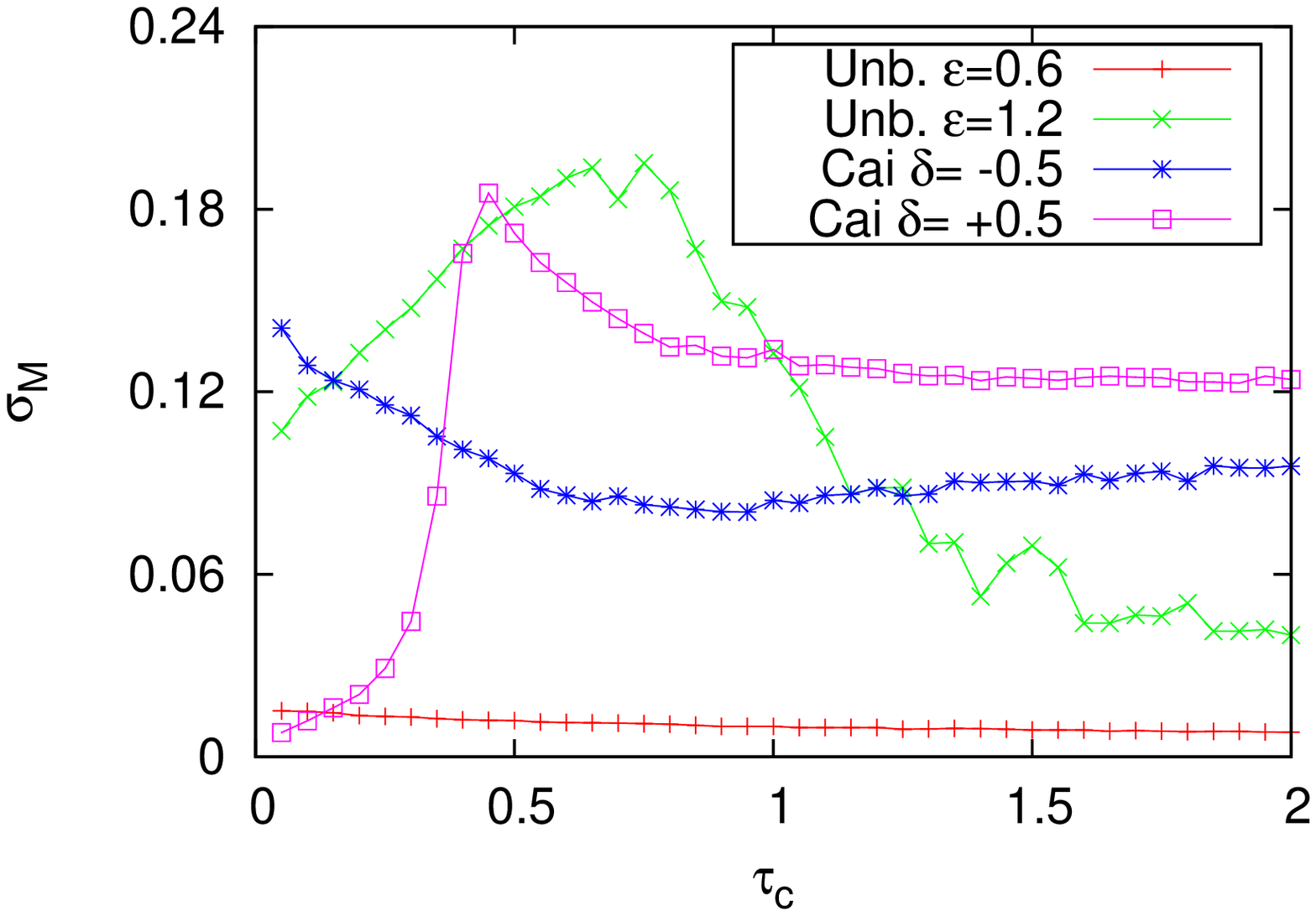}
}
\subfigure[]
{
\label{C}
\includegraphics[width=7.5cm]{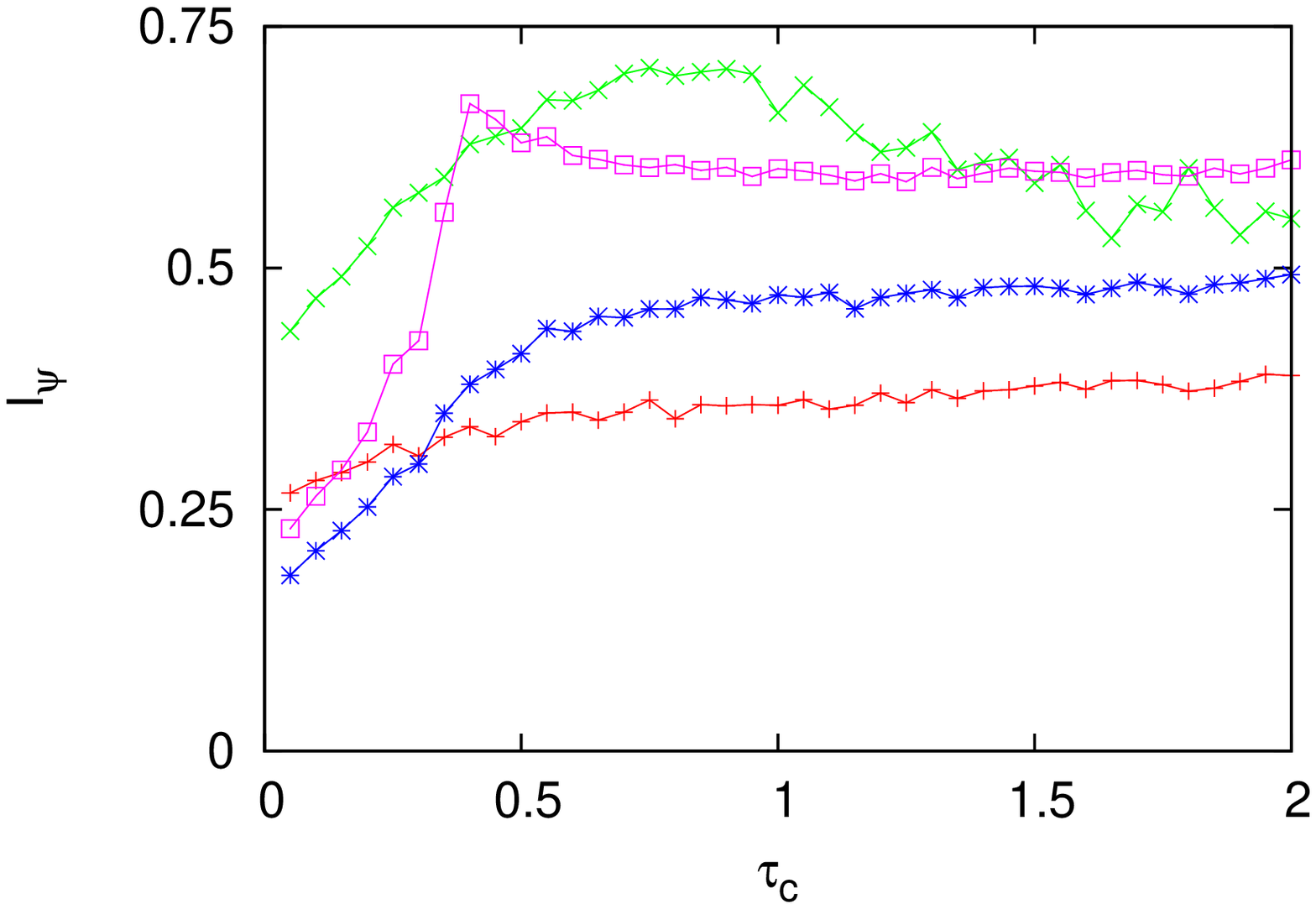}
}
\subfigure[]
{
\label{D}
\includegraphics[width=7.5cm]{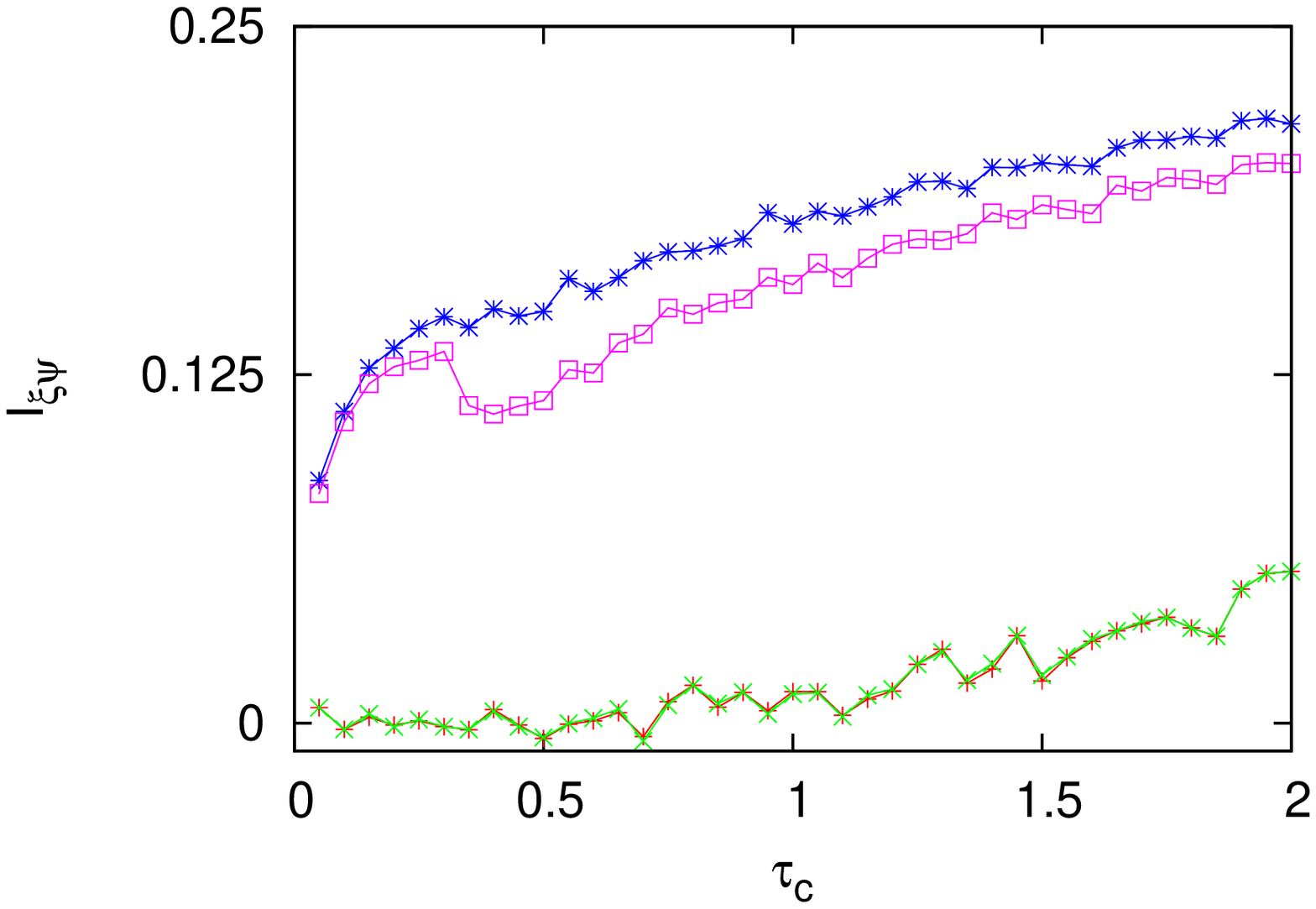}
}
\end{center}
\caption{Absence of spatial coupling in the noises ($\lambda=0$). Effects of temporal autocorrelation $\tau_c$ on GL model perturbed by additive spatially uncorrelated noises. Noise used: unbounded (GSR) and Cai with $\delta = \pm 0.5$. Top panels: global magnetization $M$ \subref{A} with its relative fluctuations $\sigma_M$ \subref{B}. Bottom panels: Moran index of field $\psi$ \subref{C} and bivariate Moran index $I_{\xi\psi}$ between the field $\left\{\psi\right\}$ and its corresponding noise one $\left\{\xi\right\}$ \subref{D}. Parameters: $40\times40$ lattice, $\lambda=0$. Cai noises, with $B=2.4$, are compared with a corresponding unbounded noise with same standard deviation, i.e. $\sqrt{2D}=B/4=0.6$ for $\delta=0.5$ and $\sqrt{2D}=B/2=1.2$ for $\delta=-0.5$.}
\label{fig_3}
\end{figure}

\begin{figure}[t]
\begin{center}
\includegraphics[width=7.cm]{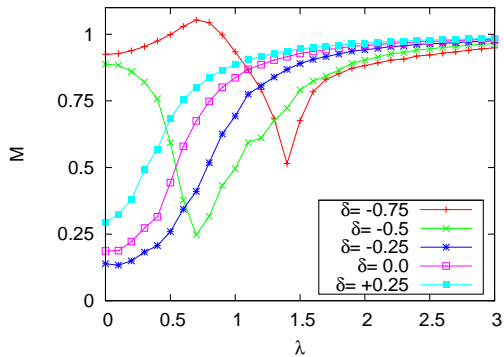}
\end{center}
\caption{Effects of Cai parameter $\delta$ on the order/disorder/order $\lambda$ transition. From the observation of $M$ one may gather that the first ordered phase is enhanced by the steepness of the bimodal distribution of the noise, which tunes both the level of magnetization and the critical value of $\lambda$. Parameters: $40\times40$ lattice, $B=2.6$ and $\tau=0.3$.}
\label{fig_4_delta}
\end{figure}

\begin{figure}[t]
\begin{center}
\subfigure[]
{
\label{A}
\includegraphics[width=7.5cm]{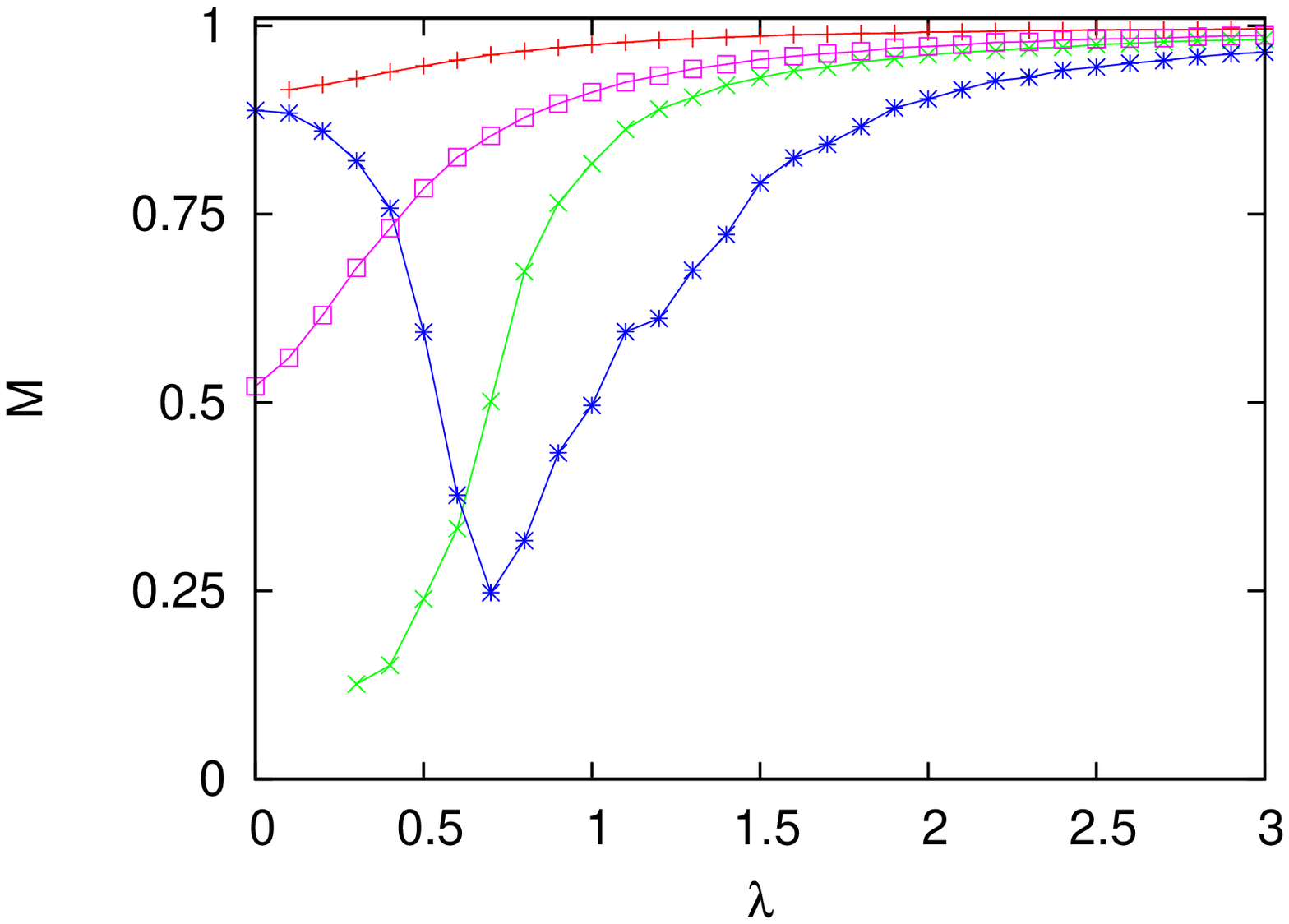}
}
\subfigure[]
{
\label{B}
\includegraphics[width=7.5cm]{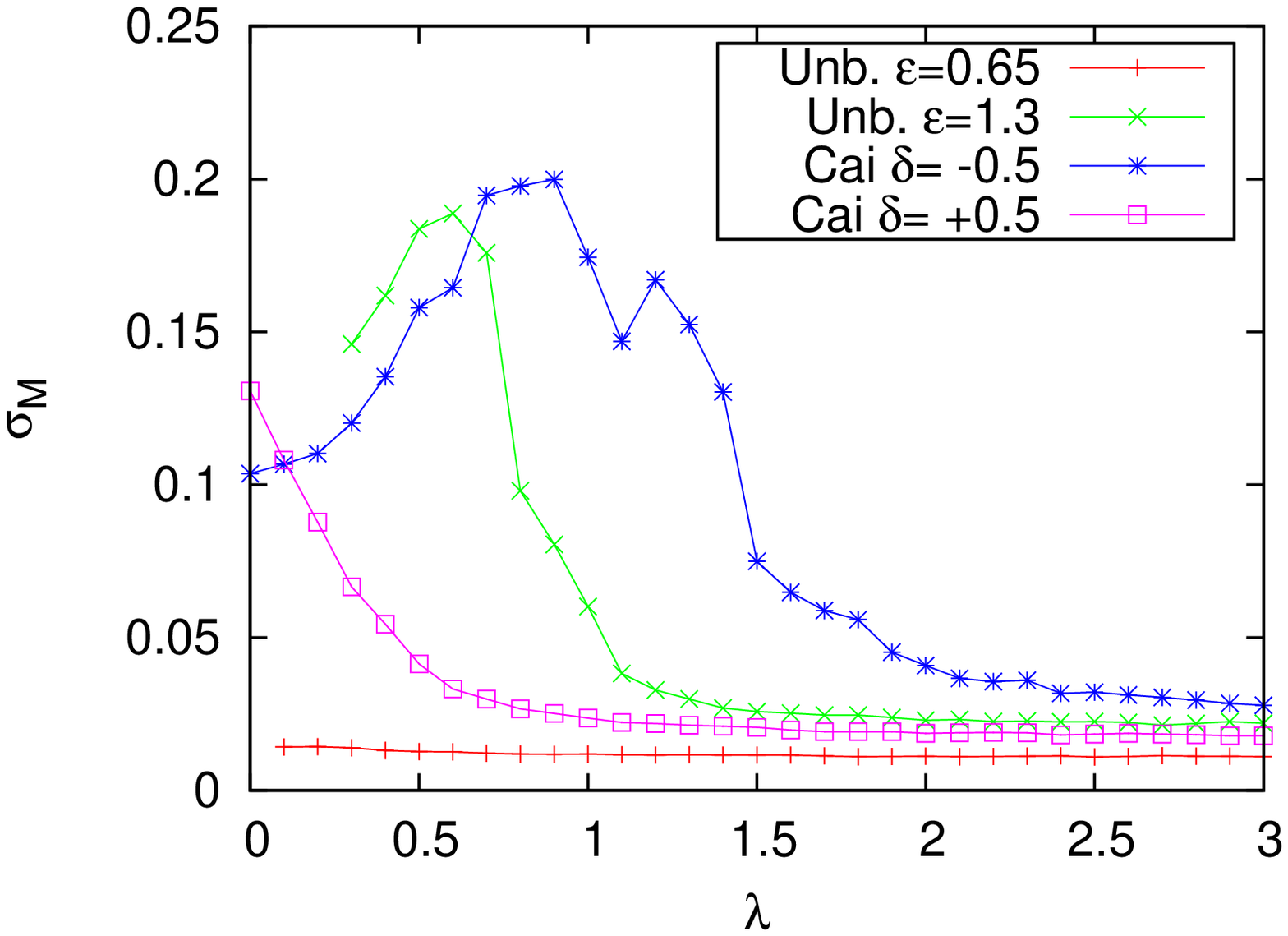}
}
\subfigure[]
{
\label{C}
\includegraphics[width=7.5cm]{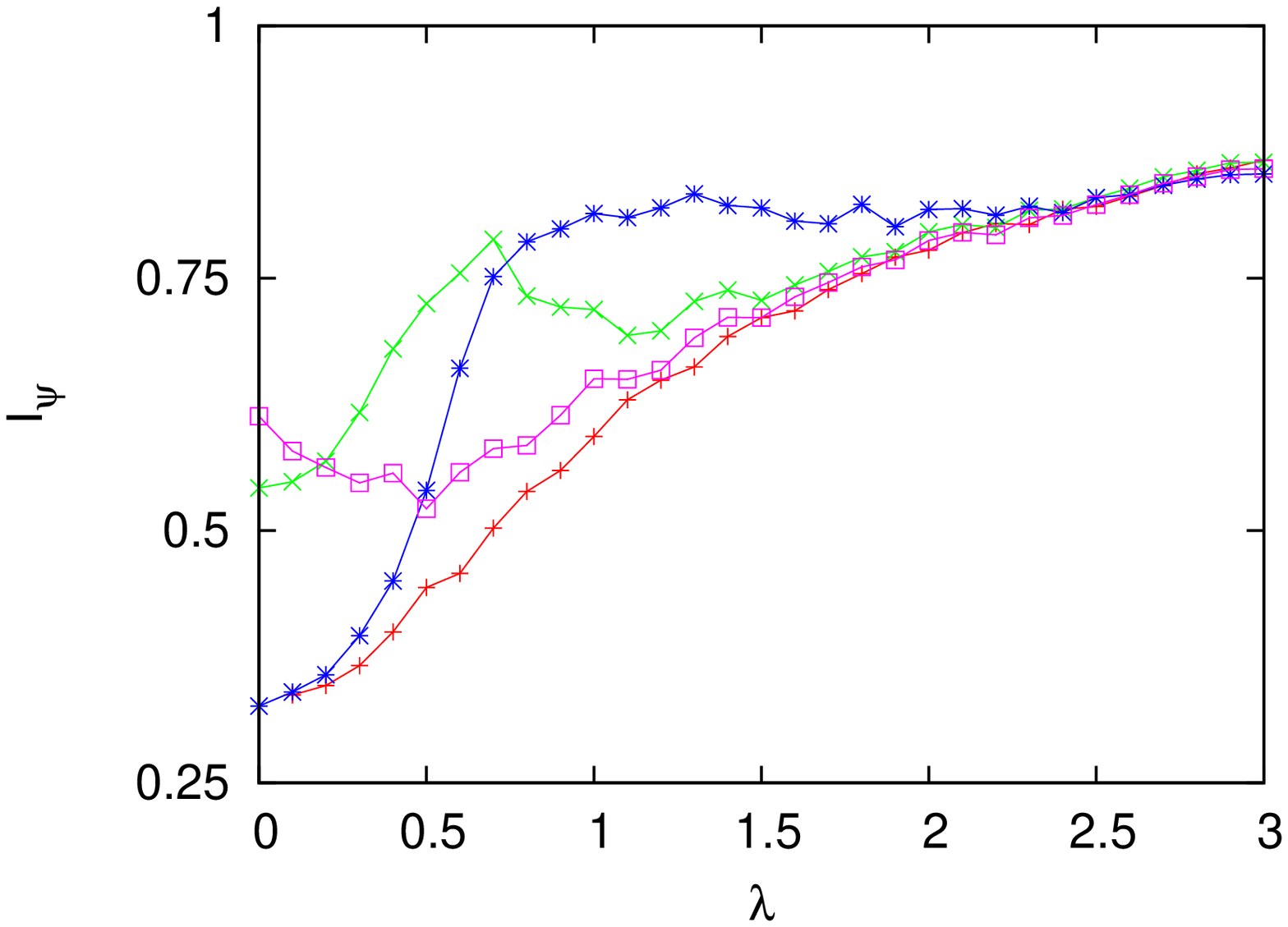}
}
\subfigure[]
{
\label{D}
\includegraphics[width=7.5cm]{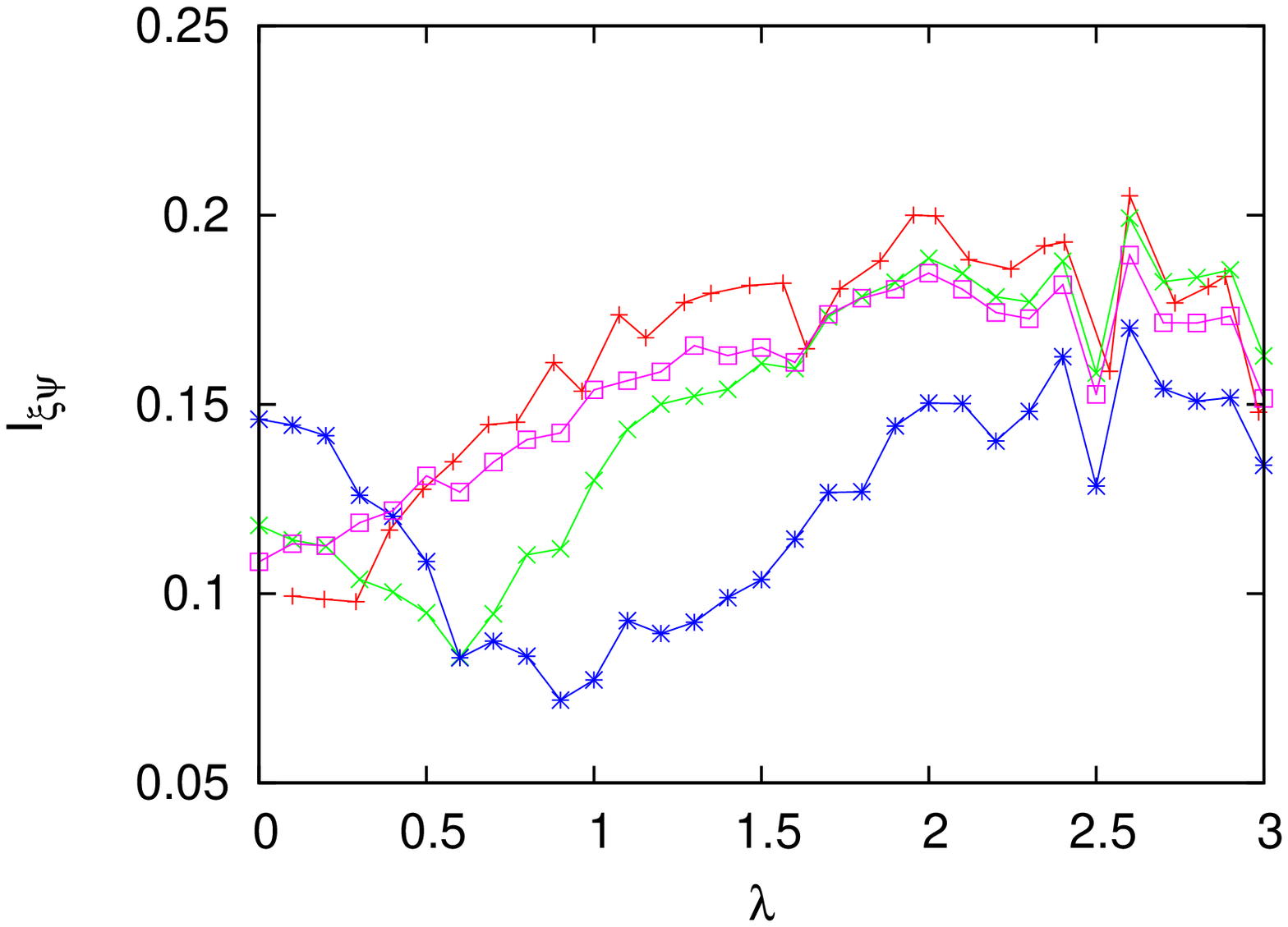}
}
\end{center}
\caption{Effects of the spatial correlation parameter $\lambda$ in case of unbounded GSR and Cai-Lin noises with $\delta = \pm 0.5$ in the additively perturbed GL model. Top panels: global magnetization $M$ (left) with its temporal variance $\sigma_M$ (right). Bottom panels: Moran index of field $\psi$ (left) and bivariate Moran index $I_{\xi\psi}$ between the field $\psi$ and its corresponding noise $\xi$ (right). Cai-Lin noises, with $B=2.6$, are compared with a corresponding unbounded noise with same standard deviation, i.e. $\sqrt{2D}=B/4=0.65$ for $\delta=0.5$ and $\sqrt{2D}=B/2=1.3$ for $\delta=-0.5$. Parameters: $40\times40$ lattice, $\tau=0.3$. Unbounded noise has $\sqrt{2D}=1.3$, while Cai noises has $B=2.6$.}
\label{fig_4}
\end{figure}

\begin{figure}[t]
\begin{center}
\includegraphics[width=7.5cm]{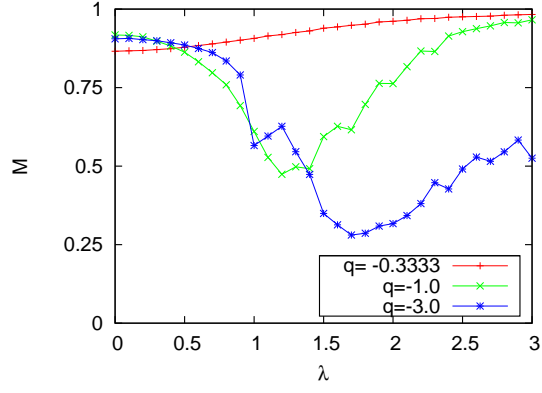}
\end{center}
\caption{Effects of Tsallis parameter $q$ on the order/disorder/order $\lambda$ transition. From the observation of $M$ one may gather that the region occupied by the second ordered phase increases with $q$, i.e. inversely with the variance of the noise distribution, which tunes both the level of magnetization and the critical value of $\lambda$. Parameters: $40\times40$ lattice, $B=2.6$ and $\tau=0.3$.}
\label{fig_4_deltaTS}
\end{figure}

In this section we compare the classical Gaussian approach to noise-induced transitions, to the new approach based on bounded noises. In particular, we focused on a well-defined case study in Gaussian noise-induced phase transitions \cite{lucafrancesco,GObook}: the real Ginzburg-Landau stochastically perturbed equation. We remark here that, although we shall show some effects of interest, here our aim is simply to illustrate the differences in the behavior of this system in response to bounded and unbounded noises.\\
In \cite{GO94,GObook} Garc\'ia-Ojalvo \textit{et al.} studied on a two dimensional lattice the real Ginzburg-Landau model:
\begin{equation}\label {GLequation}
\partial_t \psi_p= \frac{1}{2}\left(\psi_p-\psi_p^{3}+\nabla_L^2 \psi_p\right) +\xi(x,t)
\end{equation}
with external noise $\xi$ given by equation (\ref{generlattice}). The works \cite{GO94,GObook} showed that both spatial and temporal characteristic lengths of the noise $\xi(x,t)$ promote the order phase in the face of the Gaussian spatitoemporal noise $\eta$, i.e. they shift the order/disorder transition point on the right.\\
In the following we present some results from our extensive numerical study of GL equation by modeling the additive noise appearing in eq. (\ref{GLequation}) with Gaussian, Cai-Lin and Tsallis-Borland bounded noises. Note that in order to properly compare the different behaviors of the system with bounded and unbounded noise, we set:$\sigma_{unbound}=\sigma_{bound}$.\\
Similarly to the study of spatio-temporal noise described in section \ref{StuNoise}, we characterized the phase of GL model by means of the global magnetization $M=<|\sum_{(i, j)}\psi_{i,j}|>\setminus N^{2}$, of its relative fluctuations $\sigma^2_M$, as well as by using both the Moran index of the field $\psi_p$ ($I_{\psi}$) and the bivariate Moran index  between the noise field $\xi_p$ and the GL one $\psi_p$ ($I_{\xi\psi}$).\\
In our numerical study, the initial condition of GL system was the ordered phase, i.e. $\psi_p(0)=1$, thus we measured the robustness of order in the face of noise. \\
In absence of spatial coupling in the noise ($\lambda=0$), both in the Cai and in the Tsallis case the behavior of the magnetization $M$ when varying the autocorrelation time of the noise suggests that bounded noise seems to promote the disorder phase for GL system. Indeed, for bounded noises the magnetization is a decreasing function of $\tau_c$, see figure \ref{fig_3}.a (and figure 5 of SM). 
Thus temporal correlation has the opposite effect with respect to the unbounded GSR noise, which enhances the order phase\cite{GO92}. It is possible to roughly explain this effect by considering the different behavior of the equilibrium distributions $P_{eq}(\xi_p)$ in the bounded and in the unbounded noise. Indeed, in the unbounded noise the standard deviation of $\xi$ scales with $1/\tau_c$, and the transition from disordered to ordered phase with $\tau_c$ could be considered as an effect of noise amplitude reduction. On the contrary, both in Cai-Lin and Tsallis-Borland noises the equilibrium distributions do not change their standard deviation, as one can deduct by $P_{eq}(\xi)$ formulas (\ref{Peqtsallis}) and (\ref{Peqcai}), which are independent on $\tau_c$. In addition, the numerical study of the bounded noise equilibrium distribution for $\lambda\neq0$, reported in the SMs (figure 3 and 4), suggests that the influence of bounded noise in order/disorder transition would be the same also in presence of a fixed spatial correlation. 
Thus here the overall effect of $\tau_c$ consists only in increasing the spatial autocorrelation of the noise. As a consequence, the field $\psi$ is driven by an even more quenched noise, with a constant broad distribution, which enhances the disorder phase.\\
For Cai-Lin noise with $\delta>0$, the qualitative behavior of $\sigma_M$ and of the Moran index $I_{\Psi}$ are similar (see fig. \ref{fig_3}.b and \ref{fig_3}.c): they both show a maximum at transition point, for $\tau_c \approx 0.5$. In the case of Tsallis-Borland noise, $M$, $\sigma_M$ and the Moran index are similar to the ones obtained in case of Cai-Lin noise with $\delta>0$, suggesting that the GL phenomenology with temporally correlated bounded noise is similar in case of unimodal noise distribution.\\
In the case of Cai-Lin noise with bimodal distribution ($\delta<0$) the behavior of GL equation around the transition region is different (see again fig. \ref{fig_3}.b and \ref{fig_3}.c): $\sigma_M$ is a decreasing function of $\tau_c$, whereas the Moran index steadily increases with the temporal autocorrelation. Actually, our numerical results support the idea that this order/disorder transition is smoother than a second order one. In a coarse-grained description one could say that the "quasi" dichotomous feature of the equilibrium distribution allows the noise to jump back and forth between two symmetric values. Thus, in each GL field point the temporal mean value of noise is very close to zero. 
On the contrary, if the distribution is unimodal, the dispersion term slowly approach the noise to the central value $0$, so that the permanence on the same sign is longer and the temporal mean of noise in a lattice point is greater. Bimodal noise is more prone to ordered state, resulting in an increasing order in the ordered phase and in a smoother transition.\\
If one equates the variance of the bounded noise to the one of the Gaussian noise, the bound $B$ results to be larger than the standard deviation of the correspondent Gaussian noise. In spite of this, for low $\tau_c$ our two classes of bounded noises order the system better than the unbounded noise, enforcing the idea that rare value of unbounded noise, placed at the tail of distribution, could have a non-trivial role in phase ordering phenomena.\\
Finally, by computing the bivariate Moran indexes $I_{\xi\psi}$ between the discretized field $\left\{\psi_p\right\}$ and the noise $\left\{\xi_p\right\}$, we obtained that (see fig. \ref{fig_3}.d): i) unbounded noise is not correlated or, for large values of $\tau_c$, poorly correlated with the field; ii) for bounded noises, the index $I_{\xi\psi}$ is not null, and it increases with $\tau_c$, supporting the idea that the net effect of temporal correlations consists in freezing the noise, giving it quenched features. In the unbounded noise this effect is hidden by the reduction of noise amplitude, which drives the disorder/ordered transition.\\
Summarizing, in absence of spatial coupling in the bounded noise, the transition is mainly ruled by the temporal autocorrelation of the noise, i.e. by its level of "quenching". This behavior is at variance with the one corresponding to unbounded perturbations, where the transition is driven by the amplitude of noise.\\
The behavior of the system is deeply affected by the presence of spatial coupling in the noise. Note that we assessed the influence of the coupling parameter $\lambda$ in correspondence of a relatively low noise temporal autocorrelation $\tau_c=0.3$, so that the noise cannot be considered quenched-like. Indeed, the characteristic time of evolution for small $\psi$ in the deterministic case $\xi_p(t)=0$ is $2$. If the noise is of Cai-Lin type (see figure \ref{fig_4_delta} and \ref{fig_4}), a deep difference in the behavior of the magnetization in function of $\lambda$ emerges for different values of $\delta$, as illustrated in fig. \ref{fig_4_delta}. 
%\FloatBarrier
With $\delta>0$, similarly to the response to unbounded noise, there is a transition towards order, very likely due to the reduction of the variance of the distribution of the $\xi_p$'s. On the contrary, in the case $\delta = -0.5$ there is a re-entrant transition order/disorder/order. This is very likely a consequence, although non-trivial, of the stochastic bifurcation in the distribution of the $\xi_p$'s, which passes from bimodality to unimodality close to the point of minimum of $M$. For $\delta=-0.75$ the magnetization has two extremal points.\\
For $\delta = -0.5$ the spatial autocorrelation of $\psi$, measured by $I_{\psi}$, rapidly increases in that critical region; for $\delta = 0.5$ the Moran index initially decreases with $\lambda$ (which is quite counter-intuitive) and then increases (fig. \ref{fig_4}.c). Finally, the bivariate Moran index has a pattern that is on the whole increasing for $\delta = +0.5$, whereas for $\delta = -0.5$ it exhibits a minimum (fig. \ref{fig_4}.d).\\
%\FloatBarrier
In the case where the additive perturbation term in the GL equation is a Tsallis-Borland noise, one may again observe a re-entrant transition, although somewhat different from the corresponding transition found for Cai-Lin noise with same noise equilibrium distribution. This can be deduced by comparing figures \ref{fig_4_delta} and \ref{fig_4_deltaTS}, where magnetization of GL field is shown for three values of the Tsallis parameter $q$. \\
Moreover, figures \ref{fig_4_delta} and \ref{fig_4_deltaTS} suggest that the second ordered phase is enhanced by reducing the standard deviation of the noise distribution, i.e. increasing $\delta$ and $q$. This consideration supports the idea that the second ordered phase is obtained through a mechanism of noise amplitude reduction, similarly to the unbounded case. On the contrary, the first ordered phase emerges for reasons more intrinsically related to the interplay between the boundedness of the distribution and the spatio-temporal features of noise field. However, the mechanisms that allow for this phase are different in Cai-Lin and Tsallis-Borland noise, because the physics that governs their dynamics is different.\\
In all the above-examined cases, the comparisons of the different statistics made it possible to qualitatively discriminate between the two ordered phases observed in the re-entrant transitions. Indeed, the first one, which appears at small $\lambda$, is characterized by low values of $I_\psi$ (blue curve in fig. \ref{fig_4}.c, for $\lambda<0.5$), and in it order is enhanced by the low temporal correlation of the noise. On the contrary, the second ordered phase, appearing at larger $\lambda$, is characterized by large values of $I_\psi$ (fig. \ref{fig_4}.c, for $\lambda>1.5$), and in it the order is caused by both the lower amplitude of the noise and by the strong spatial correlation of the noise.
\begin{figure}[t]
\centering
\subfigure[]
{
\label{A}
\includegraphics[scale=0.3, angle=-90]{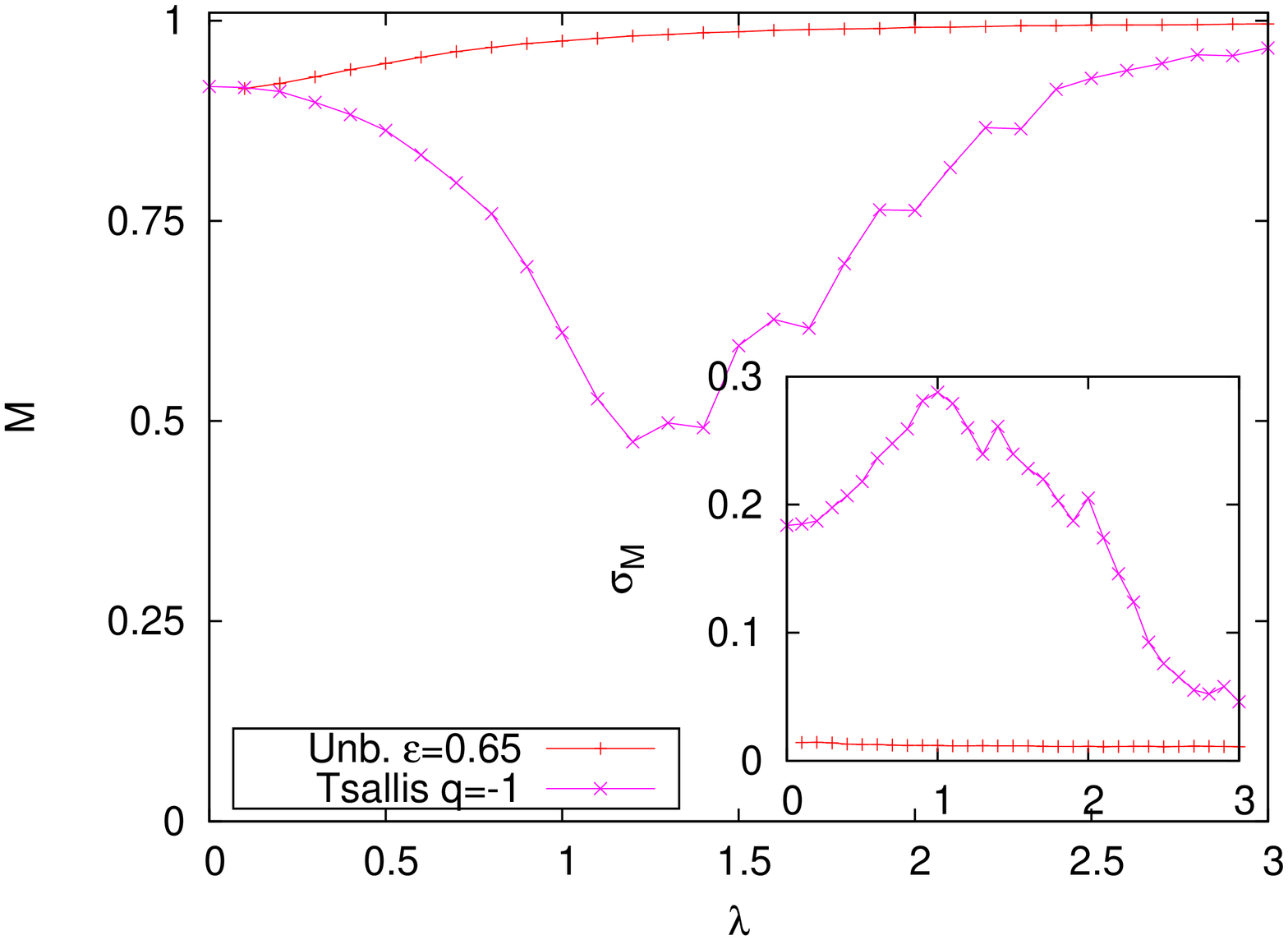}
}
%\subfigure[]
%{
%\label{B}
%\includegraphics[width=7.5cm]{GLfig4Tsallisvar.eps}
%}
\subfigure[]
{
\label{C}
\includegraphics[scale=0.3, angle=-90]{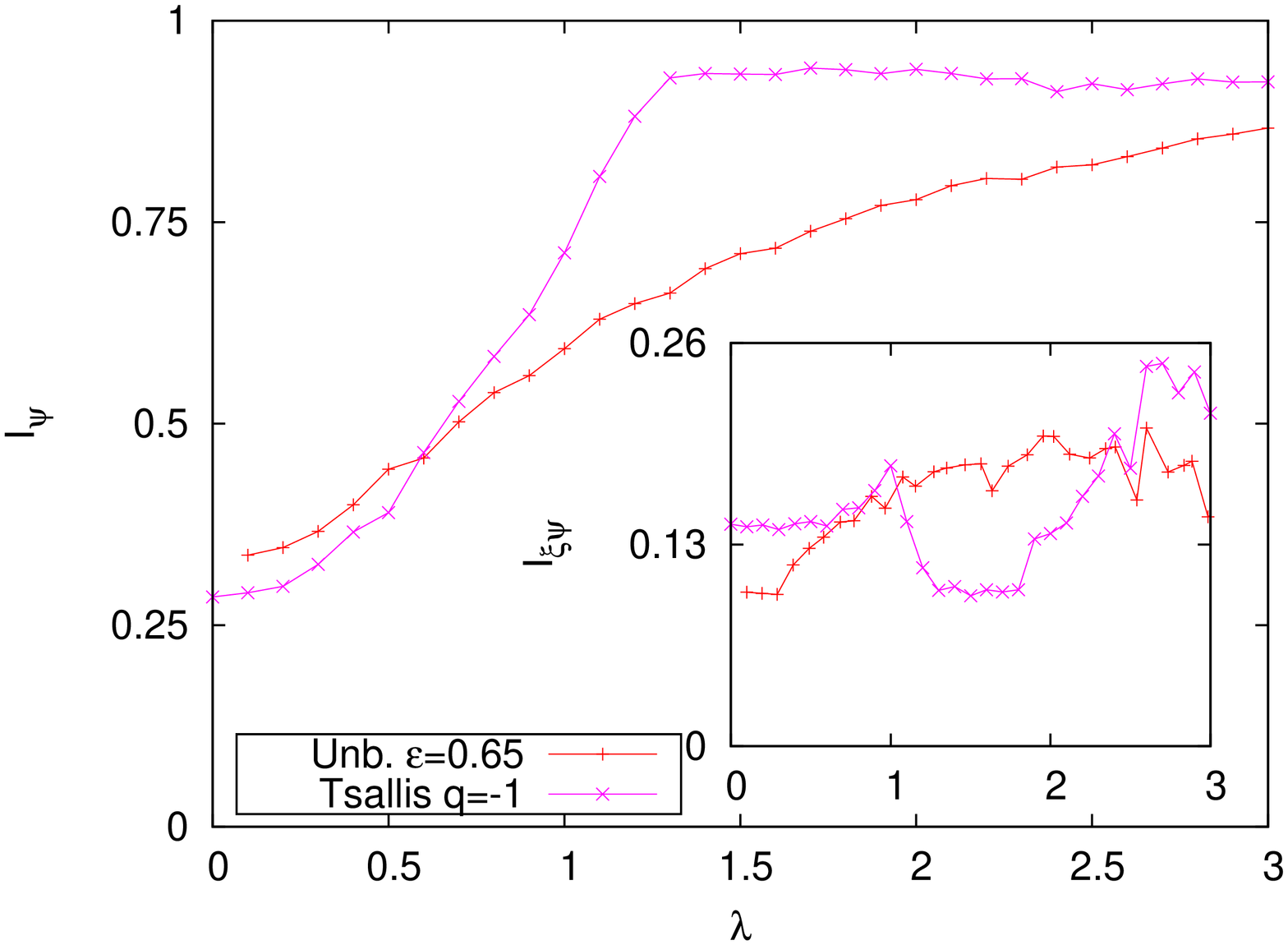}
}
%\subfigure[]
%{
%\label{D}
%\includegraphics[width=7.5cm]{GLfig4TsallismoranBi.eps}
%}
\caption{Effects of the spatial correlation parameter $\lambda$ in case of unbounded and Tsallis-Borland noises with $q=-1$ in the additively perturbed GL model. Panel \subref{A}: global magnetization $M$ with relative fluctuations $\sigma_M$. Panel \subref{C}: Moran index of field $\psi$ and bivariate Moran index $I_{\xi\psi}$ between the field $\left\{\psi\right\}$ and its corresponding noise one $\left\{\xi\right\}$. Tsallis noise, with $B=2.6$, is compared with an unbounded case with same standard deviation, i.e. $\sqrt{2D}=B/4=0.65$. With bounded noise, GL system shows an double order/disorder/order transition.}
\label{fig_4_Tsallis}
\end{figure}
%\FloatBarrier
Finally, in figure \ref{BTRANS} we show the effect of the amplitude $B$ on the distribution of the GL field. Increasing $B$, for non-small values of $\lambda$ one moves from unimodal density mainly located close to $\psi=1$ to a bimodal density.
\begin{figure}[htb!]
\begin{center}
\includegraphics[width=7.5cm]{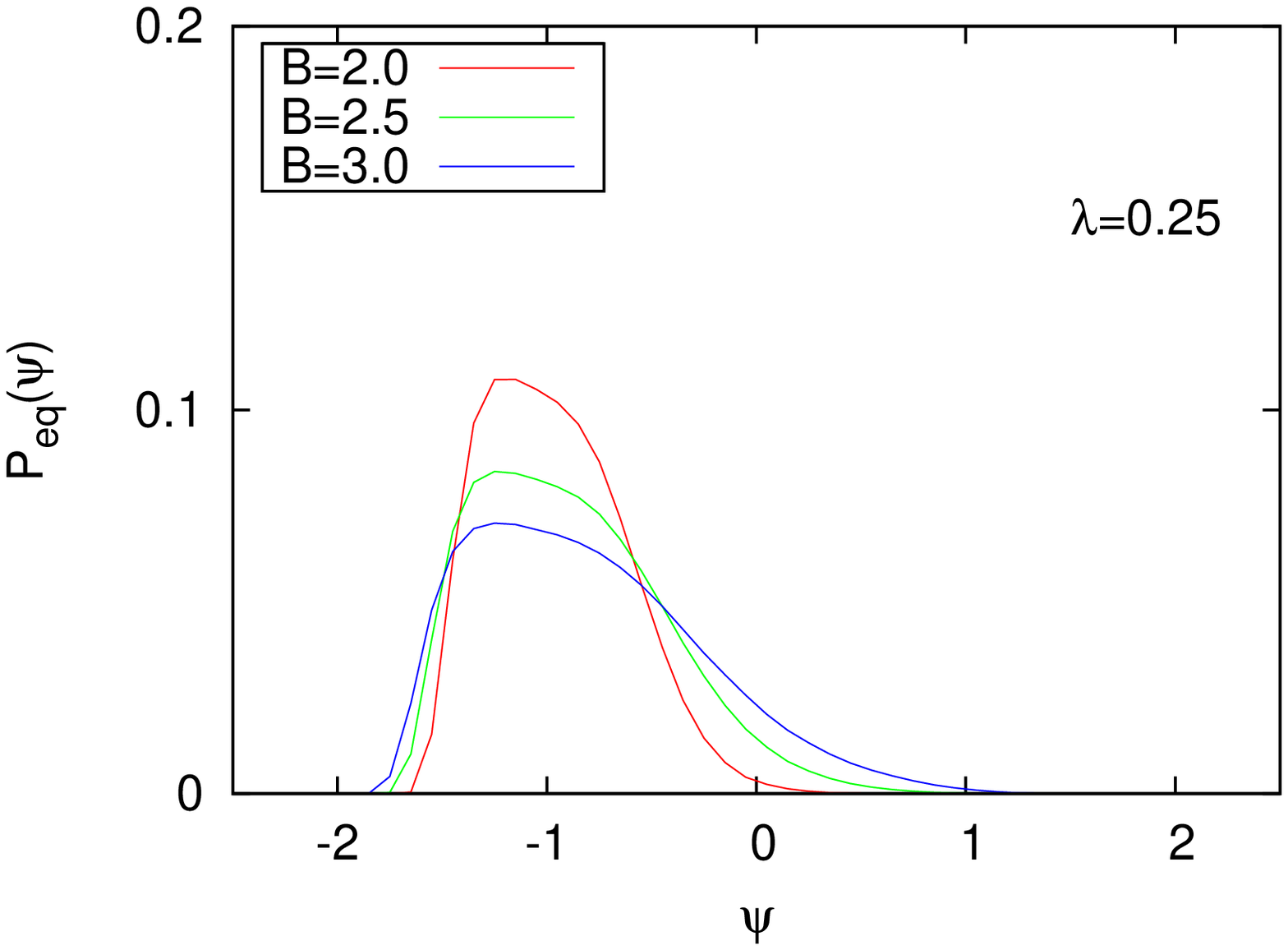}
\includegraphics[width=7.5cm]{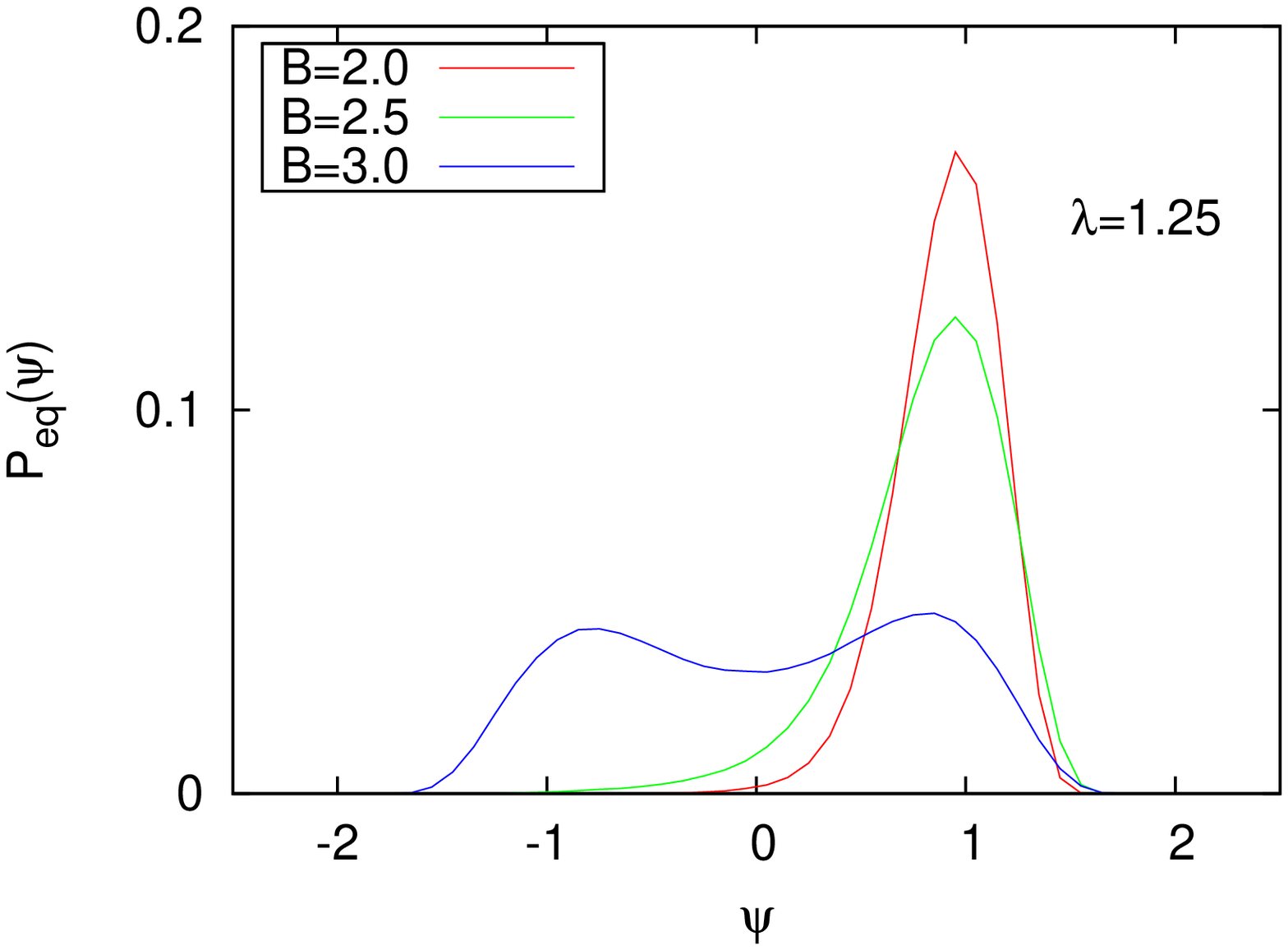}
\end{center}
\caption{ Equilibrium distribution $P_{eq}(\psi_p)$ of GL system on a $40\times40$ lattice, for Cai noise with different sizes of bound. Noise parameters are $\delta=-0.5$ and $\tau=0.3$.}
\label{BTRANS}
\end{figure} 
\section{Concluding remarks}
In the first part of this work, we defined two classes of spatio-temporal colored bounded noises, which can be directly derived by two zero-dimensional bounded noises: the Tsallis noise and the Cai noise. We have analyzed the role of the spatial coupling parameter $\lambda$ and of the temporal correlation parameter $\tau$ on the distribution of the noise by employing suitable statistical observables. Differently from unbounded noise case, the equilibrium distribution of the studied bounded noises does not depend on $\tau$, while in some cases the increase of $\lambda$ induces transitions from bimodality to unimodality in the distribution. These features could be relevant when bounded noises are applied to dynamical systems, in particular in the presence of noise-induced phase transitions.\\
With this in mind, in the second part we employed the above-mentioned two kinds of bounded noises to investigate the phase transitions of the Ginzburg-Landau model under additive stochastic perturbations. \\ Our simulations showed a phenomenology quite different from the one induced by colored unbounded noises.\\ 
To start, in presence of spatially uncoupled bounded noises, the increase of the temporal correlations enhances the quenching of the noise, eventually producing an "order-to-disorder" transition in the GL model. Note that when the perturbation is unbounded an opposite transition is observed.\\ 
Furthermore, spatial coupling induces contrasting effects on the spatio-temporal fluctuations of the noise, resulting for some kind of noises in a re-entrant transition (order/disorder/order) in the response of the GL system. This specific case of noise type-dependence in spatio-temporal dynamical systems is novel, and is in line with previous observations in zero-dimensional systems \cite{dongan, pre}.\\
Finally, we stress that in defining the spatial extensions of the Cai-Lin and of the Tsallis-Borland temporal noises we adopted as spatial coupling the classical Laplace operator, in line with \cite{GO92}. However, more sophisticated alternatives might be employed, to take into account long-range interplays \cite{lucafrancesco, murray} provided that the constraint of boundedness of the noise can be preserved. 
\bibliographystyle{unsrt} 
\bibliography{Bibliography} 
\end{document}

%7 \preprint{APS/123-QED}

\title{Supplementary materials of the manuscript 'Spatiotemporal Bounded Noises, and transitions induced by them in solutions of real Ginzburg-Landau model'}% Force line breaks with \\

\author{Sebastiano de Franciscis}
\author{Alberto d'Onofrio}

\date{\today}%
\maketitle
\section{A generalization of the Tsallis-Borland bounded noise}
The Tsallis-Borland bounded noise noise is here generalized as follows:
\begin{equation}\label{tbwExtended}
\xi^{\prime} (t)= \frac{1 }{\tau} f(\xi) + \frac{\sqrt {2 D} }{\tau}\eta(t),
\end{equation}
where $f(+B)=-\infty \ f(-B)=+\infty$. Our extension of the Tsallis-Borland noise allows to generate a noise with a pre-assigned stationary density $P_*$. Indeed, from the Fokker-Planck equation associated to eq. (\ref{tbwExtended}), one immediately obtains that the following $f(\xi)$ must be chosen to be :
\begin{equation}\label {simple}
f(\xi)= \frac{D}{\tau^2} \frac{P_*^{\prime}(\xi)}{P_*(\xi)}.
\end{equation}
For example, in order to obtain the density $P(\xi)=(1/2)Cos(\xi)_+$, one must consider:
$$ f(\xi)= -\frac{D}{\tau} tan(\xi).$$
\section{Useful Transformations}
Note that in Tsallis noise the unboundedness of $f(\xi)$ at $\xi=|B|$ may cause problematic stiffness during numerical simulations. For example in the Euler schema
$$ \xi(t + \Delta t)= \xi(t) + \frac{1 }{\tau} f(\xi) \Delta t + \frac{\sqrt {2 D} }{\tau} G_t \sqrt {\Delta t} $$
if $\xi(t)$ is close to $|B|$ then it may be $ |\xi(t + \Delta t)| > B $.\\ 
A possible remedy, suggested for the Tsallis-Borland noise in ref. 20 of the main text (H. S. Wio and R. Toral. Physica D, 193, 161-168, 2004) is to re-extract the gaussian random variable $G_t$.\\
Another trick, which we introduce here, consists in applying the following nonlinear transformation to 
\begin{equation}\label{noiseSINtrans}
 \xi = B \ sin(\phi),  
\end{equation}
which yields the following Langevin equation for $\phi(t)$:
\begin{equation}\label {LangTransf}
\phi^{\prime} = a(\phi) + b(\phi)\eta(t),
\end{equation}
where:
\begin{equation}\label {bgeneralNOSpace}
b(\phi)= \frac{\sqrt{2 D}}{\tau B}\frac{1}{cos(\phi)}
\end{equation}
\begin{equation}\label {ageneralNOSpace}
a(\phi)= \frac{D}{B^2 \tau^2} \frac{sin(\phi)}{cos^3(\phi)}+\frac{f(B \ sin(\phi))}{B \ cos(\phi)}.
\end{equation}
Thus, in the case of temporal Tsallis noise (equation (1) in the text) we have:
\begin{equation}\label {aNOSpace}
a(\phi)= -\frac{1+q}{2\tau} \frac{sin(\phi)}{cos^3(\phi)}
\end{equation}
\begin{equation}\label {bNOSpace}
b(\phi)= \sqrt{ \frac{2}{\tau_c} \left(\frac{1-q}{5-3 q} \right)}\frac{1}{cos(\phi)}
\end{equation}
In the spatiotemporal case, similarly to the purely temporal case it is useful to define the transformation:
$$ \xi_p = B \ sin(\phi_p) $$
leading to the $N$ equations:
\begin{equation}
\phi_p^{\prime} (t)= a_p(\phi)+b_p(\phi)\eta_p(t).
\end{equation}
where:
\begin{equation}\label {bgeneralSpace}
b_p(\phi)= \frac{\sqrt{2 D}}{\tau B}\frac{1}{cos(\phi_p)}
\end{equation}
\begin{equation}\label {ageneralSpace}
a_p(\phi)= \frac{D}{B^2 \tau^2} \frac{sin(\phi_p)}{cos^3(\phi_p)}+\frac{f(B \ sin(\phi_p))}{B \ cos(\phi_p)} +\frac{\lambda^2}{2 \tau_c}\frac{1}{cos(\phi_p)}h^{-2}\sum_{n \in Ne(p)}\left(sin(\phi_n)-sin(\phi_p)\right).
\end{equation}
Namely, in the case of spatiotemporal Tsallis noise (equation (11) in the text) we have:
\begin{equation}\label {aSpace}
a_p(\phi)= -\frac{(1+q)}{2\tau} \frac{sin(\phi_p)}{cos^3(\phi_p)}+\frac{\lambda^2}{2 \tau_c}\frac{1}{cos(\phi_p)}h^{-2}\sum_{n \in Ne(p)}\left(sin(\phi_n)-sin(\phi_p)\right)
\end{equation}
\begin{equation}\label {bSpace}
b_p(\phi)= \sqrt{ \frac{2}{\tau_c} \left(\frac{1-q}{5-3 q} \right)}\frac{1}{cos(\phi_p)}.
\end{equation}
Similarly to the Tsallis' noise, also in simulations of the Cai noise one may, although less frequently, encounter numerical stiffness. By applying the transformation:
\begin{equation}\label{trasf}
\xi = B \ sin(\phi) 
\end{equation}
one gets a stochastic equation of the form of eq. (\ref{LangTransf}), where:
\begin{equation}\label {CAIbgeneralNOSpace}
b(\phi)= \frac{g(B \ sin(\phi))}{B \ cos(\phi)}
\end{equation}
\begin{equation}\label {CAIageneralNOSpace}
a(\phi)=tan(\phi)\left(\frac{1}{2 B^2 cos^2(\phi)}-\frac{1}{\tau_C} \right).
\end{equation}
In the case of temporal Cai-Lin noise (equation (3) in the text) we have:
\begin{equation}\label {CAIaNOSpace}
a(\phi)= -\frac{1+2 \delta}{2(1+\delta)\tau_c} tan(\phi)
\end{equation}
\begin{equation}\label {CAIbNOSpace}
b(\phi)= \sqrt{ \frac{1}{\tau_c(1+\delta)}}Signum(Cos(\phi)) 
\end{equation}
When simulating the spatiotemproal Cai-Lin noise, applying the transformation (\ref{trasf}) yields the following $N$ stochastic equations:
$$\phi_p^{\prime} (t)= a_p(\phi)+b_p(\phi)\eta_p(t),$$
where:
\begin{equation}\label {latticeCAIbgeneralNOSpace}
b_p(\phi)= \frac{g(B \ sin(\phi_p))}{B \ cos(\phi_p)}
\end{equation}
\begin{equation}\label {latticeCAIageneralSpace}
a_p(\phi)=tan(\phi_p)\left(\frac{1}{2 B^2 cos^2(\phi_p)}-\frac{1}{\tau_C} \right) + \frac{\lambda^2}{2 \tau_c}\frac{1}{cos(\phi_p)}h^{-2}\sum_{n \in Ne(p)}\left(sin(\phi_n)-sin(\phi_p)\right).
\end{equation}
Namely, in case of the spatiotemporal Cai-Lin noise (equation (12) in the text) we have:
\begin{equation}\label {latticeCAIaSpace}
a(\phi)= -\frac{1+2 \delta}{2(1+\delta)\tau_c} tan(\phi_p)+\frac{\lambda^2}{2 \tau_c}\frac{1}{cos(\phi_p)}h^{-2}\sum_{n \in Ne(p)}\left(sin(\phi_n)-sin(\phi_p)\right)
\end{equation}
\begin{equation}\label {latticeCAIbNOSpace}
b_p(\phi)= \sqrt{ \frac{1}{\tau_c(1+\delta)}}Signum(Cos(\phi_p)).
\end{equation}

\section{Numerical Algorithms}
Both Cai and Tsallis spatiotemporal noise generative equations, once reduced to the form (\ref{noiseSINtrans}), by mean of the trasformation (\ref{LangTransf}), can be considered has a Langevine equation in $\phi$ with multiplicative noise. We have integrated this noise numerically with the derivative free Milstein method for It\^{o} rapresentation, i.e.:
\begin{eqnarray*}
\tilde{\phi}_{n}&=&\phi_n+a_n dt+b_n \sqrt{dt} \eta_n \\
\phi_{n+1}&=&\phi_n+a_n dt+b_n \sqrt{dt} \eta_n+\frac{\sqrt{dt}}{2}\left[b\left(\tilde{\phi}_n\right)-b_n\right]\left[\eta_n^{2}-1\right].
\end{eqnarray*}
Once noise values $\xi=Bsin\left(\phi(t)\right)$ have been obtained, GL equation is integrated  with a standard Runge-Kutta method:
\begin{eqnarray*}
\partial_t \psi_p &=& F\left(\vec{\psi}\right) +\xi_p\\
\tilde{\psi_p}     &=&\psi_p(t)+\frac{dt}{2}\left(F\left(\vec{\psi}\right)+\xi_p\right)\\
\tilde{\psi_p}     &=&\psi_p(t)+\frac{dt}{4}\left(F\left(\vec{\psi}\right)+F\left(\vec{\tilde{\psi}}\right)\right)+dt\xi_p.
\end{eqnarray*}
We have used a time step $dt=0.005$, for a total time of $T=250$, taking the mean values of the observables in the last interval of lenght $25$, whit a number of different runs between $10$ and $20$.  

\section{Supplementary Figures}

\begin{figure}[htb!]
\begin{center} 
\subfigure[]
{
\label{C}
\includegraphics[width=7.5cm]{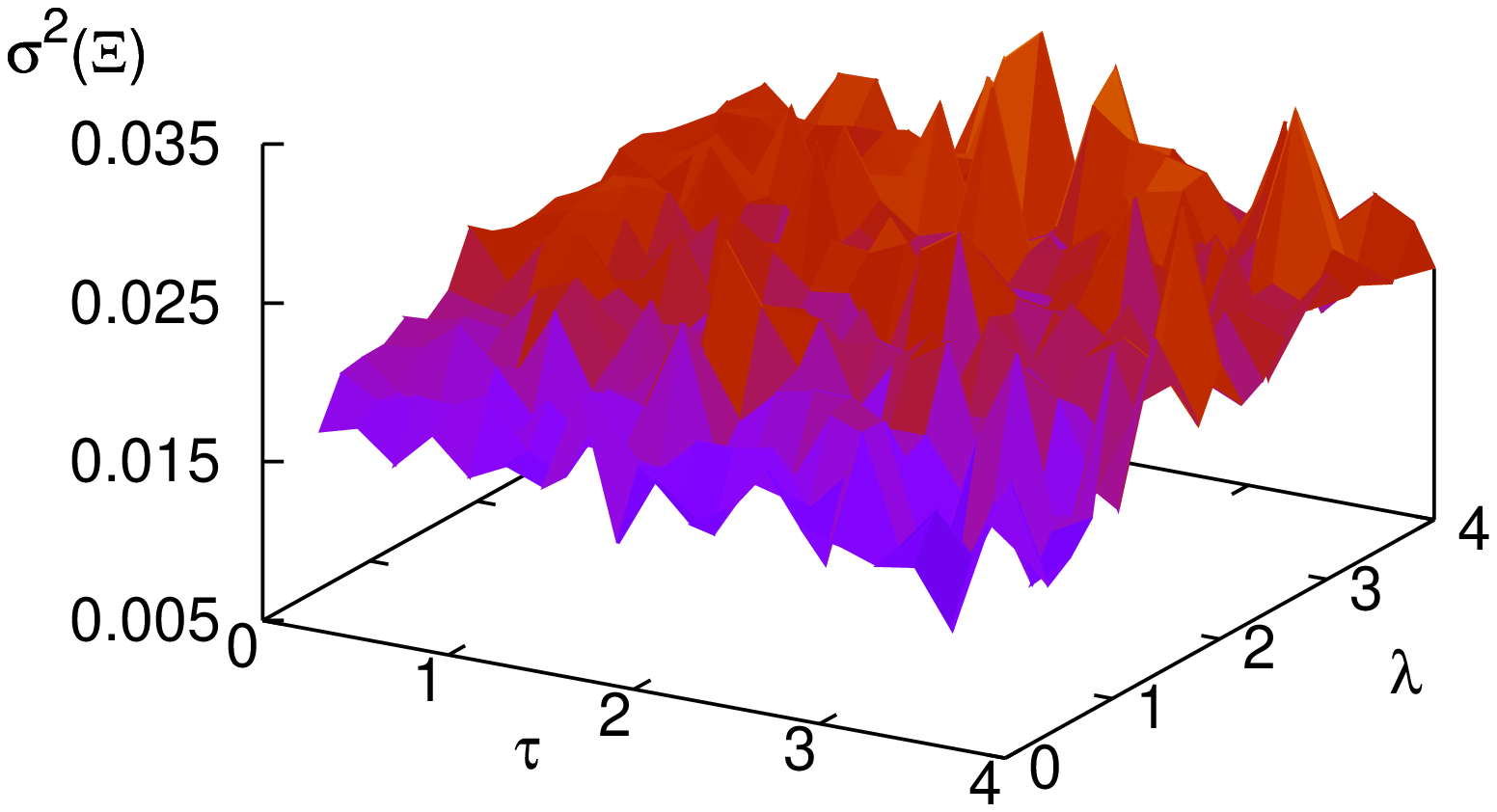}
}
\subfigure[]
{
\label{D}
\includegraphics[width=7.5cm]{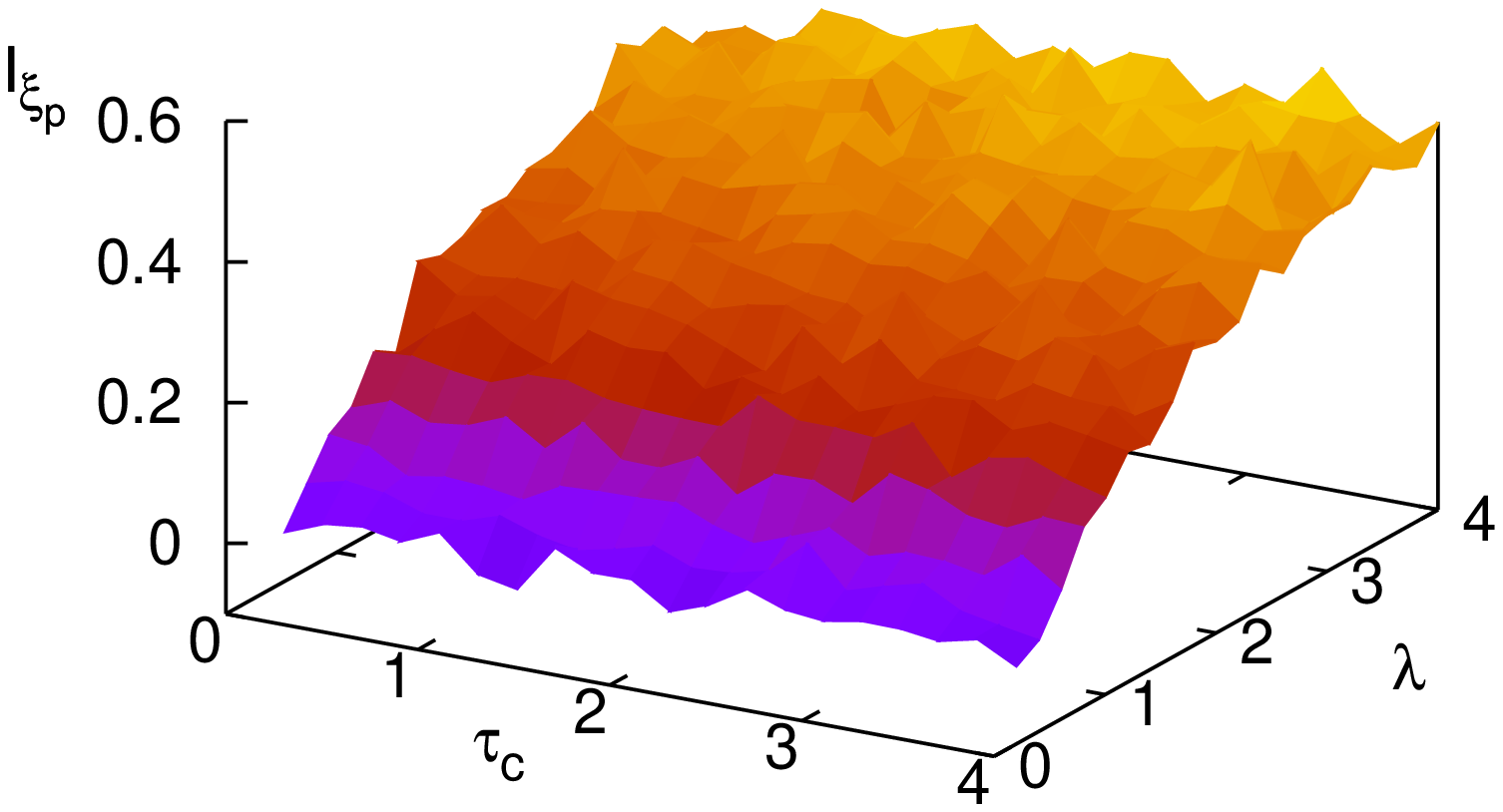}
}
\end{center}
\caption{Characterization of spatiotemporal fluctuations for bounded Cai noise: plots of $\sigma\left(\Xi\right)$ \subref{C} and Moran index \subref{D}. Parameters: $40\times40$ lattice, $B=1$ and $\delta=-0.5$.}
\label{fig_Surf}
\end{figure}

\begin{figure}[htb!]
\begin{center} 
\subfigure[]
{
\label{A}
\includegraphics[width=7cm]{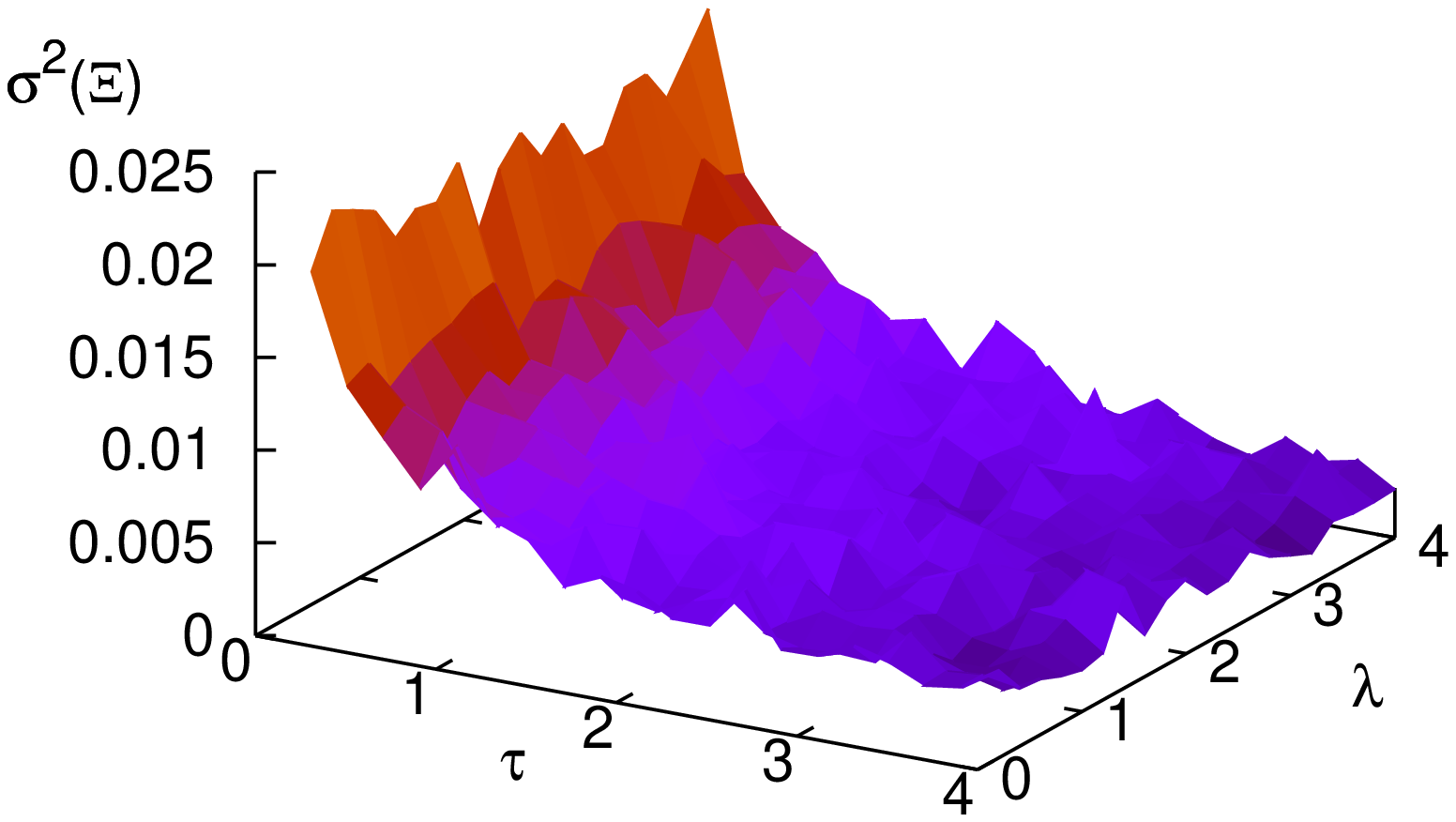}
}
\subfigure[]
{
\label{B}
\includegraphics[width=7cm]{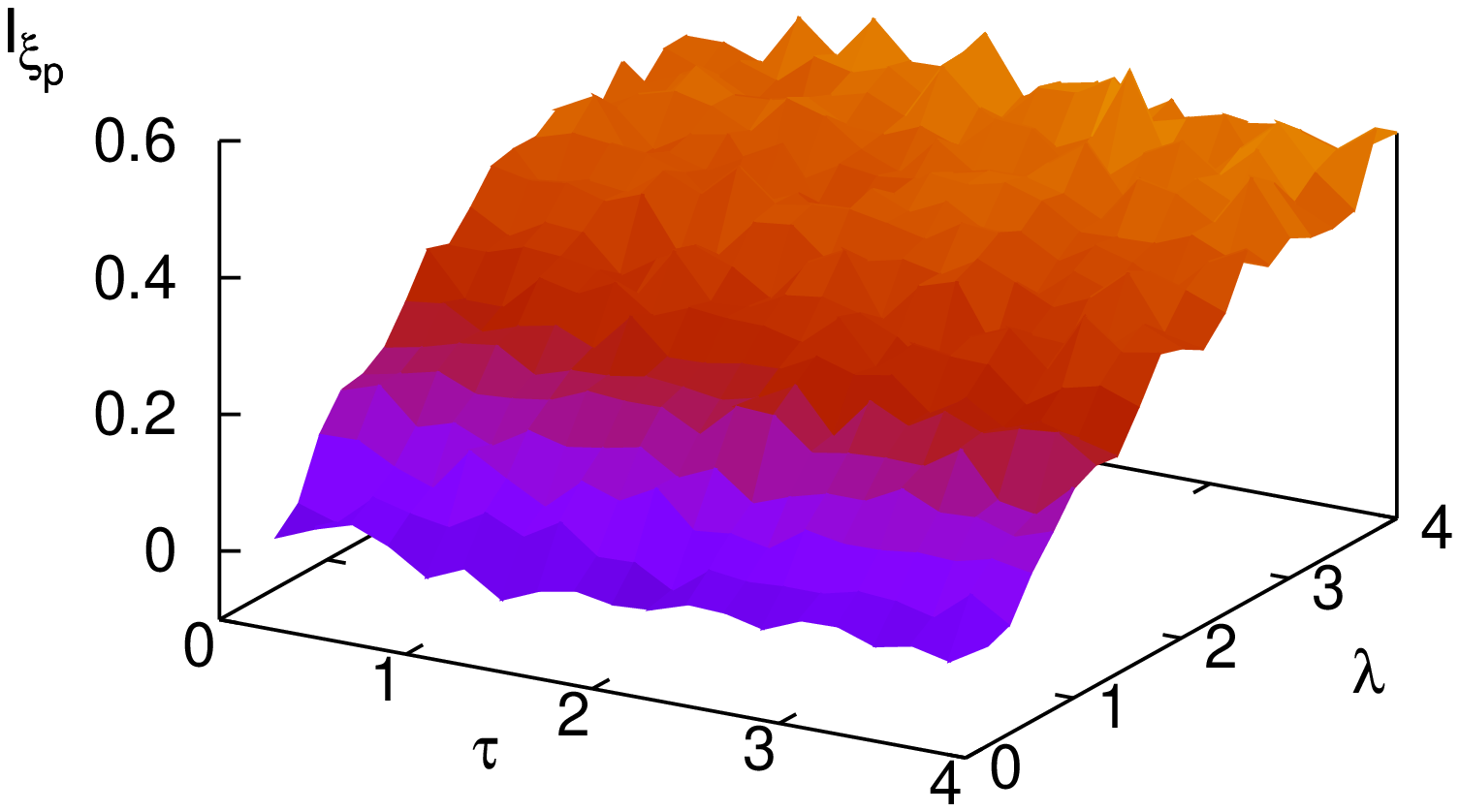}
}
\end{center}
\caption{Characterization of spatiotemporal fluctuation for unbounded spatiotemporal noise. Plots of $\sigma\left(\Xi\right)$ \subref{A} and Moran index \subref{B}. Parameters: $40\times40$ lattice, $\sqrt{2D}=0.5$.}
\label{fig_GOeps0p5}
\end{figure}

\begin{figure}[h]
\begin{center}
\subfigure[]
{
\label{A}
\includegraphics[width=7cm]{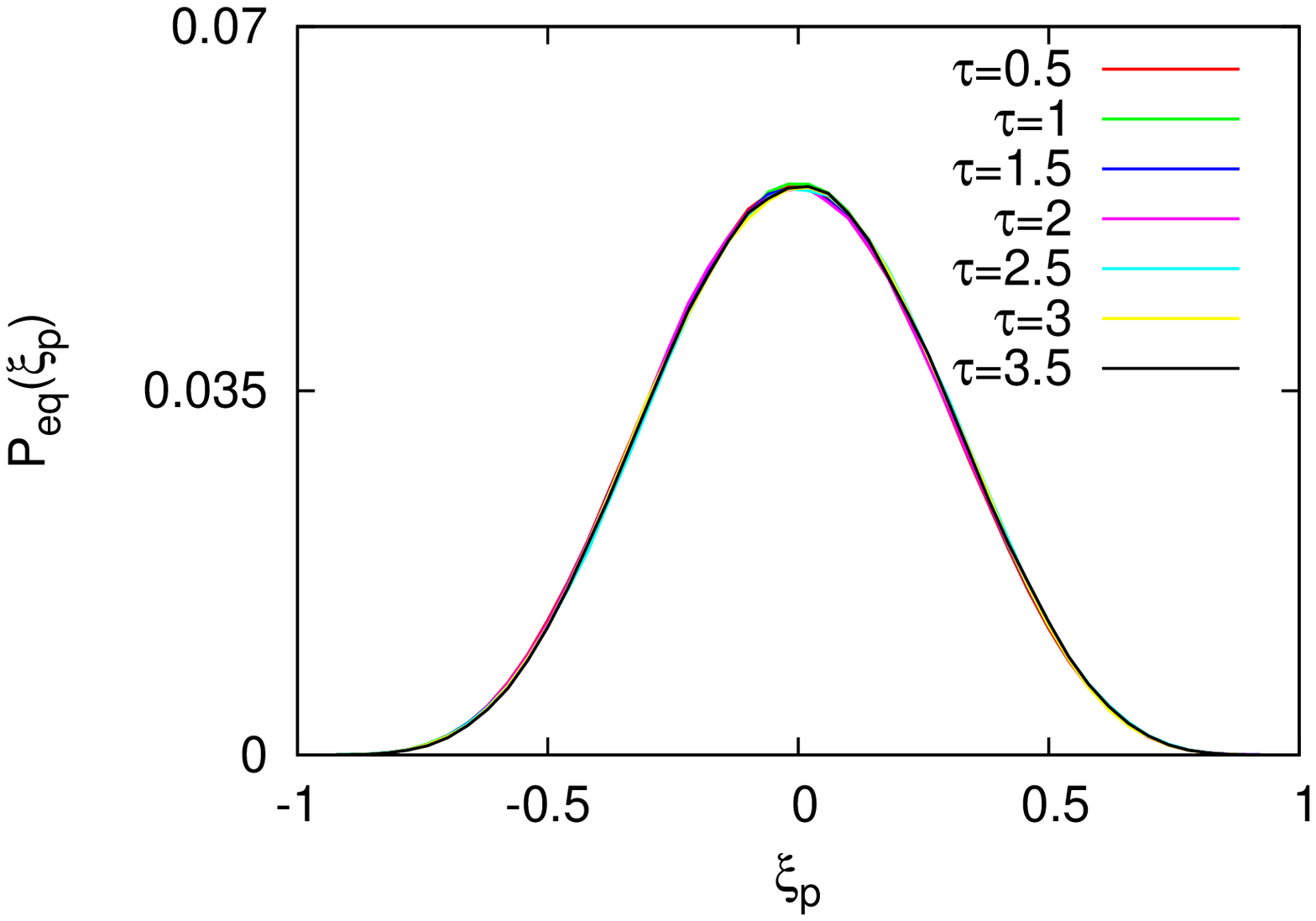}
}
\subfigure[]
{
\label{B}
\includegraphics[width=7cm]{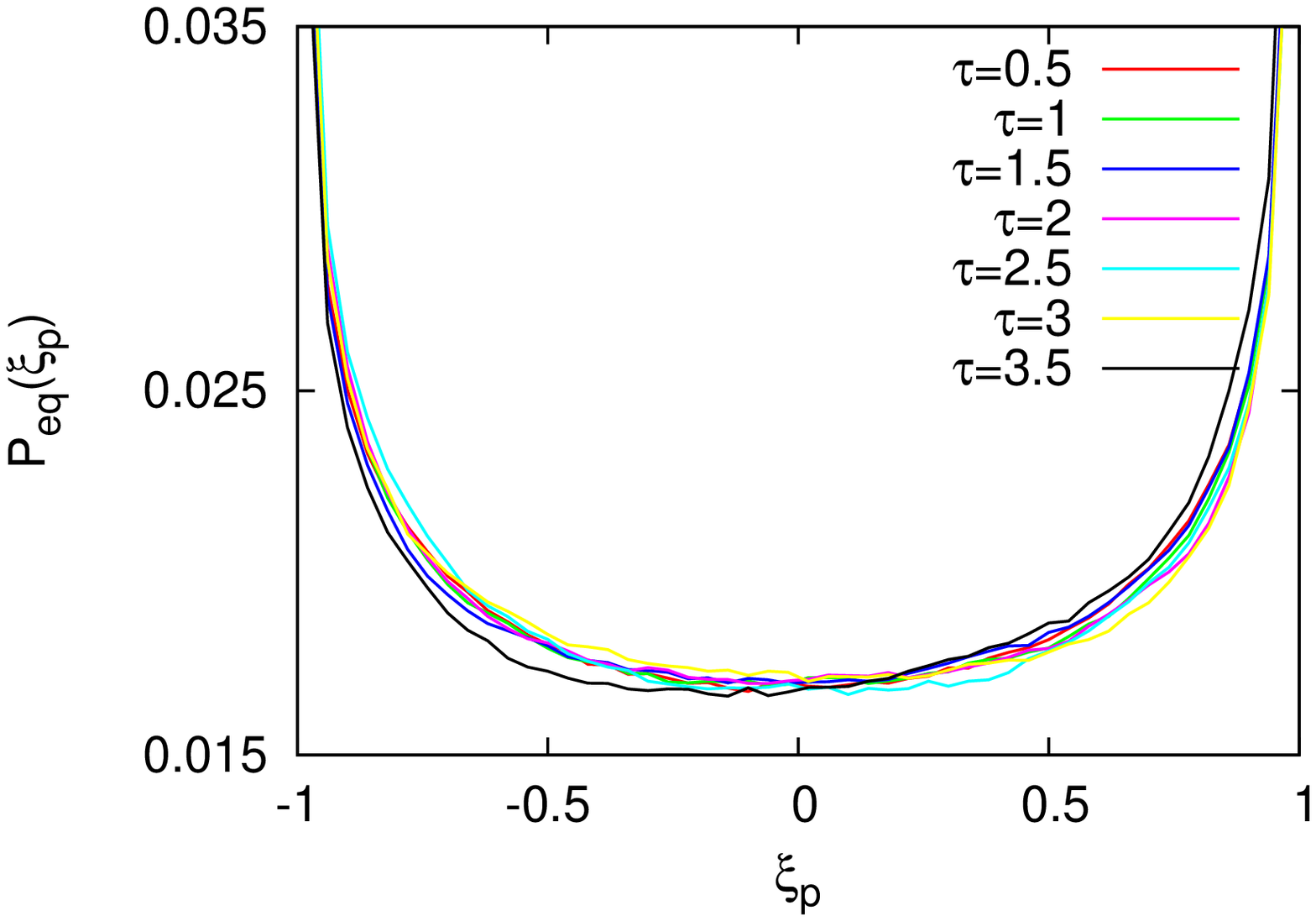}
}
\end{center}
\caption{Influence of characterisctic correlation length $\tau$ on Cai bounded noise equilibrium distribution $P_{eq}(\xi_p)$ of a $40\times40$ lattice system. Panel \subref{A}: Cai noise with $B=1$, $\lambda=1$ and $\delta=+0.5$. Panel \subref{B}: same noise with $\lambda=0.5$ and $\delta=-0.5$.}
\label{fig_P_eq_lambdaconst}
\end{figure}
\begin{figure}[h]
\begin{center}
\subfigure[]
{
\label{A}
\includegraphics[width=7.5cm]{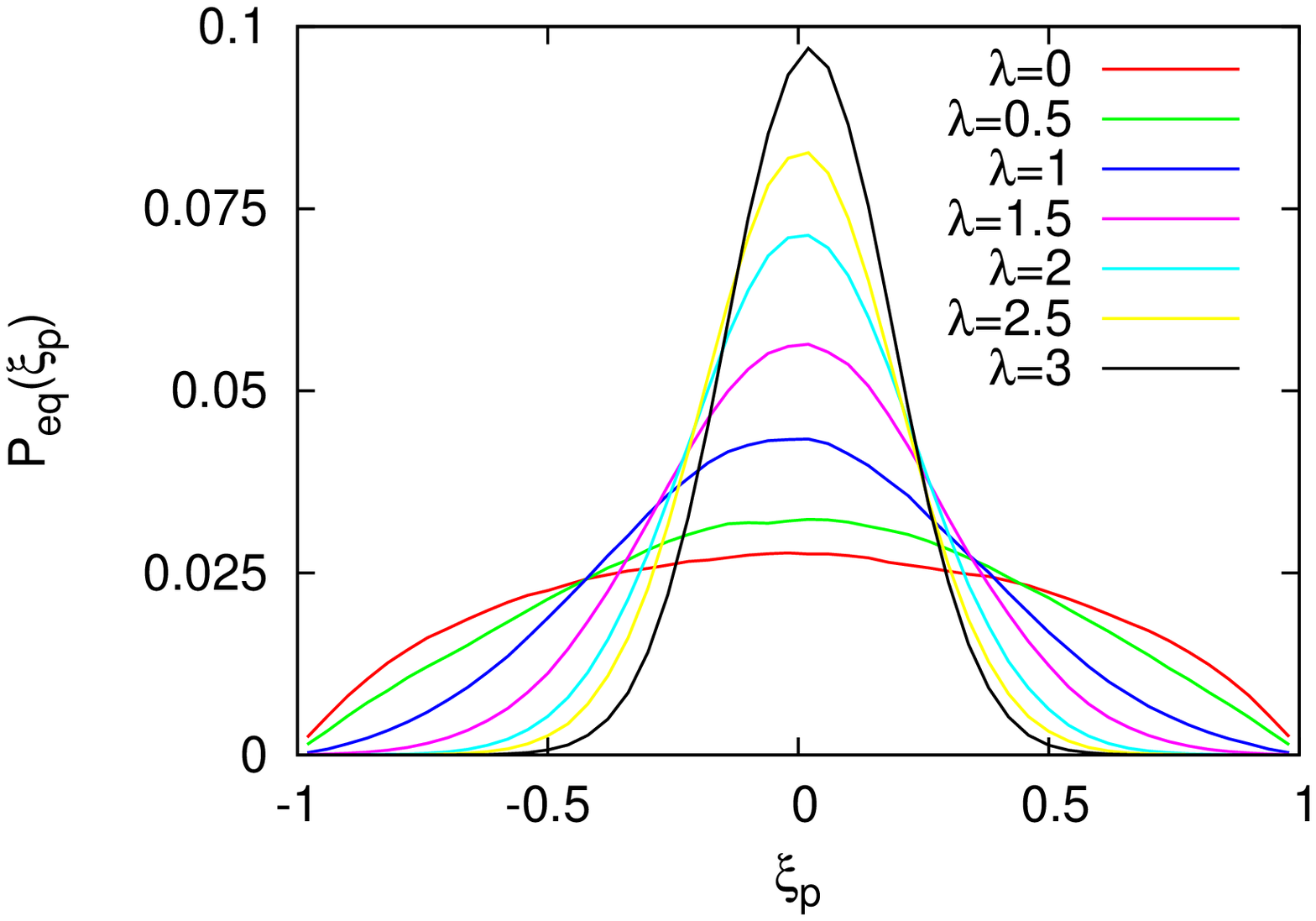}
}
\subfigure[]
{
\label{B}
\includegraphics[width=7.5cm]{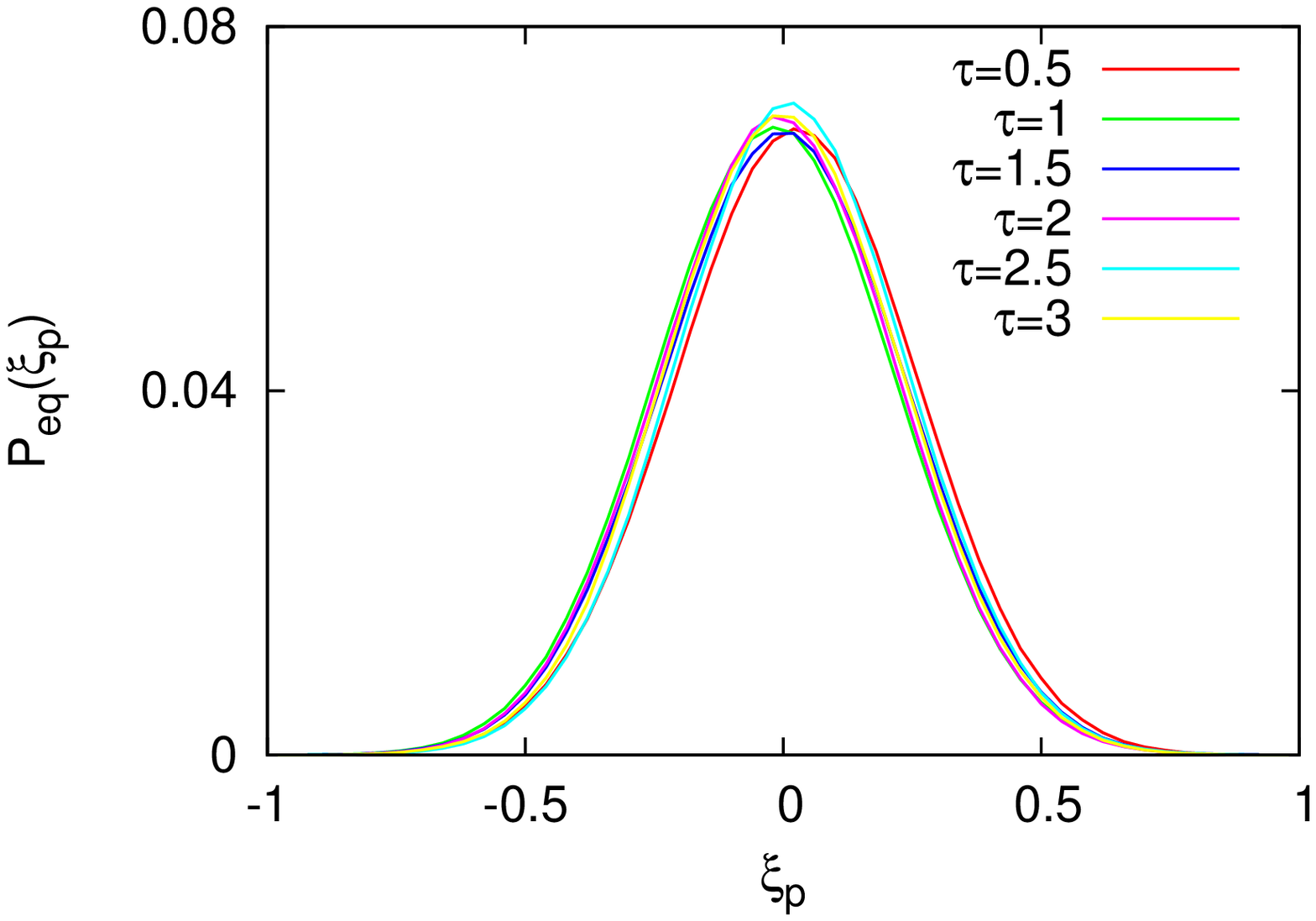}
}
\end{center}
\caption{Influence of $\tau$ and $\lambda$ on Tsallis bounded noise equilibrium distribution $P_{eq}(\xi_p)$ of a $40\times40$ lattice system. Parameters are $B=1$,  $q=-1$, $\tau=2$ for panel \subref{A} and $\lambda=2$ for panel \subref{B}.}
\label{fig_P_eq_Tsallis}
\end{figure}
\begin{figure}[t]
\begin{center}
\includegraphics[width=7cm]{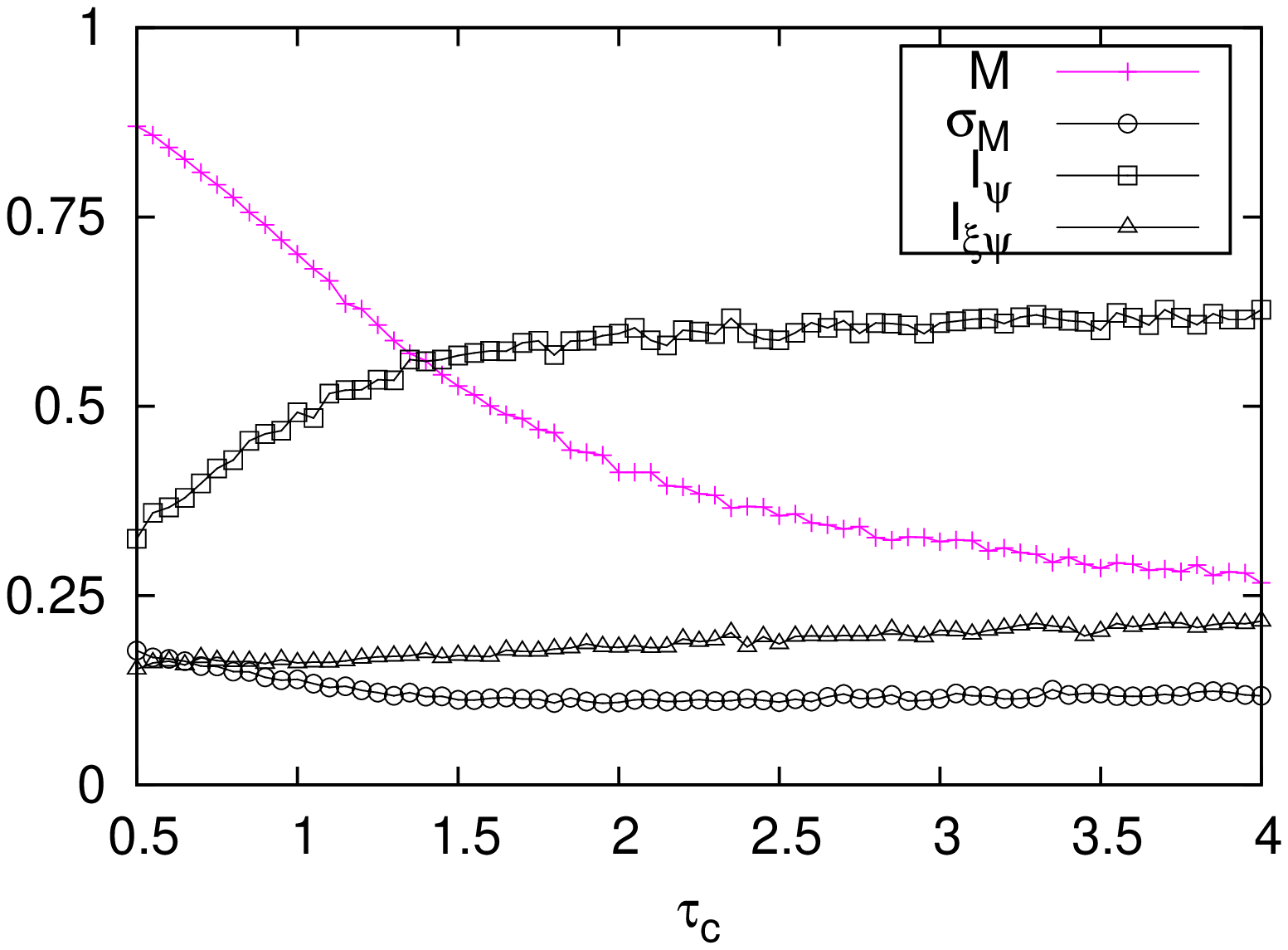}
\end{center}
\caption{Effects of temporal autocorrelation $\tau$ on a discretized GL model in a $40\times40$ lattice perturbed by Tsallis noise, with $q=-1$ and $B=2.4$ and $\lambda=0$. GL phases are characterized by the global magnetization $M$, the relative fluctuations $\sigma_M$, the Moran index of field $\psi$ and the bivariate Moran index $I_{\xi\psi}$ between the field $\left\{\psi\right\}$ and its corresponding noise one $\left\{\xi\right\}$.
%Bounded noise can be compared with the unbounded case with same standard deviation, i.e. $\sqrt{2D}=B/4=0.6$, depicted in figure \ref{fig_3}. 
With bounded noise, GL system shows an inverted order/disorder transition with respect to the unbounded case.} 
\label{fig_3_Tsallis}
\end{figure}
\begin{figure}[b]
\begin{center}
\subfigure[]
{
\label{A}
\includegraphics[width=7cm]{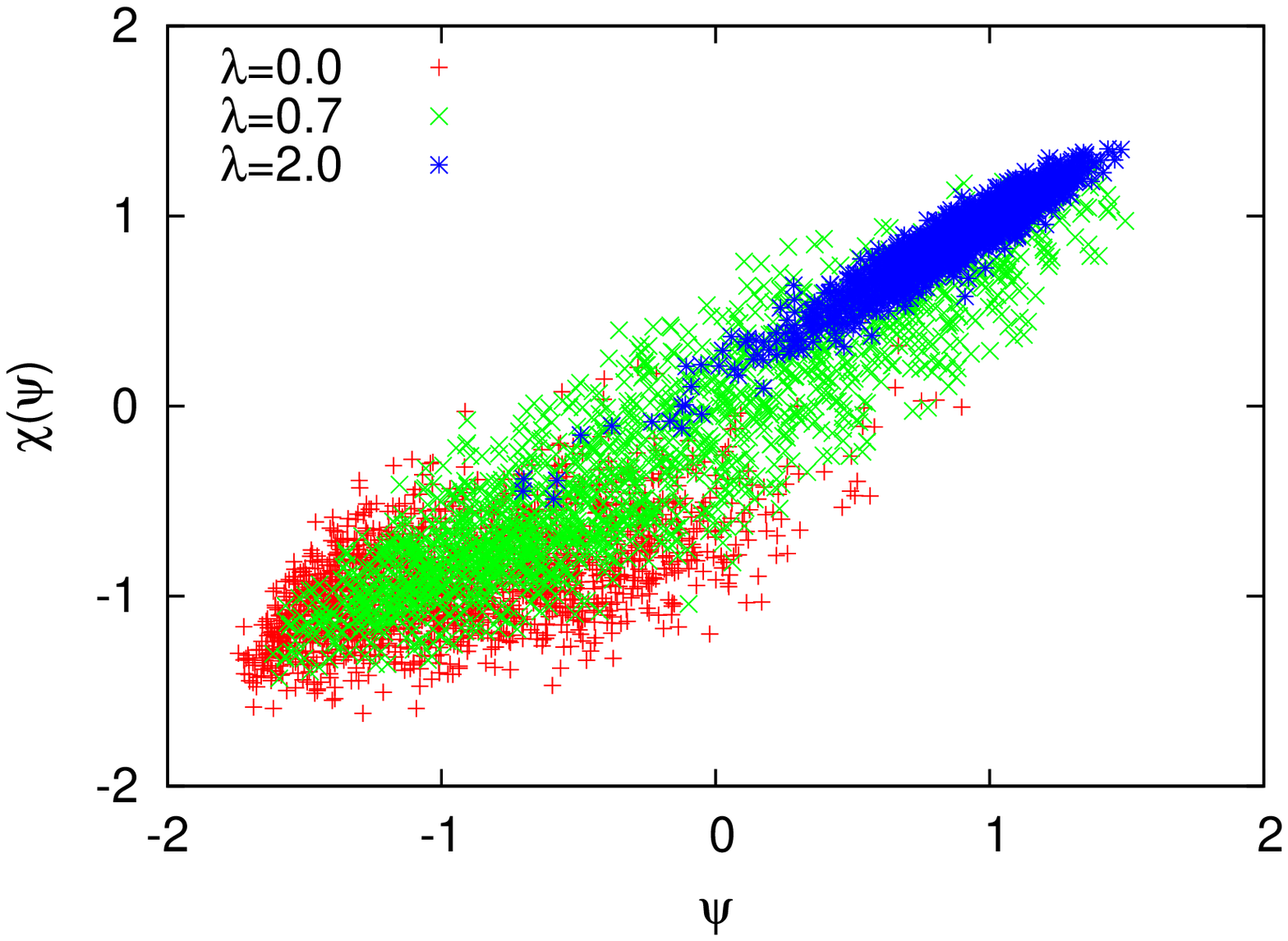}
}
\subfigure[]
{
\label{B}
\includegraphics[width=7cm]{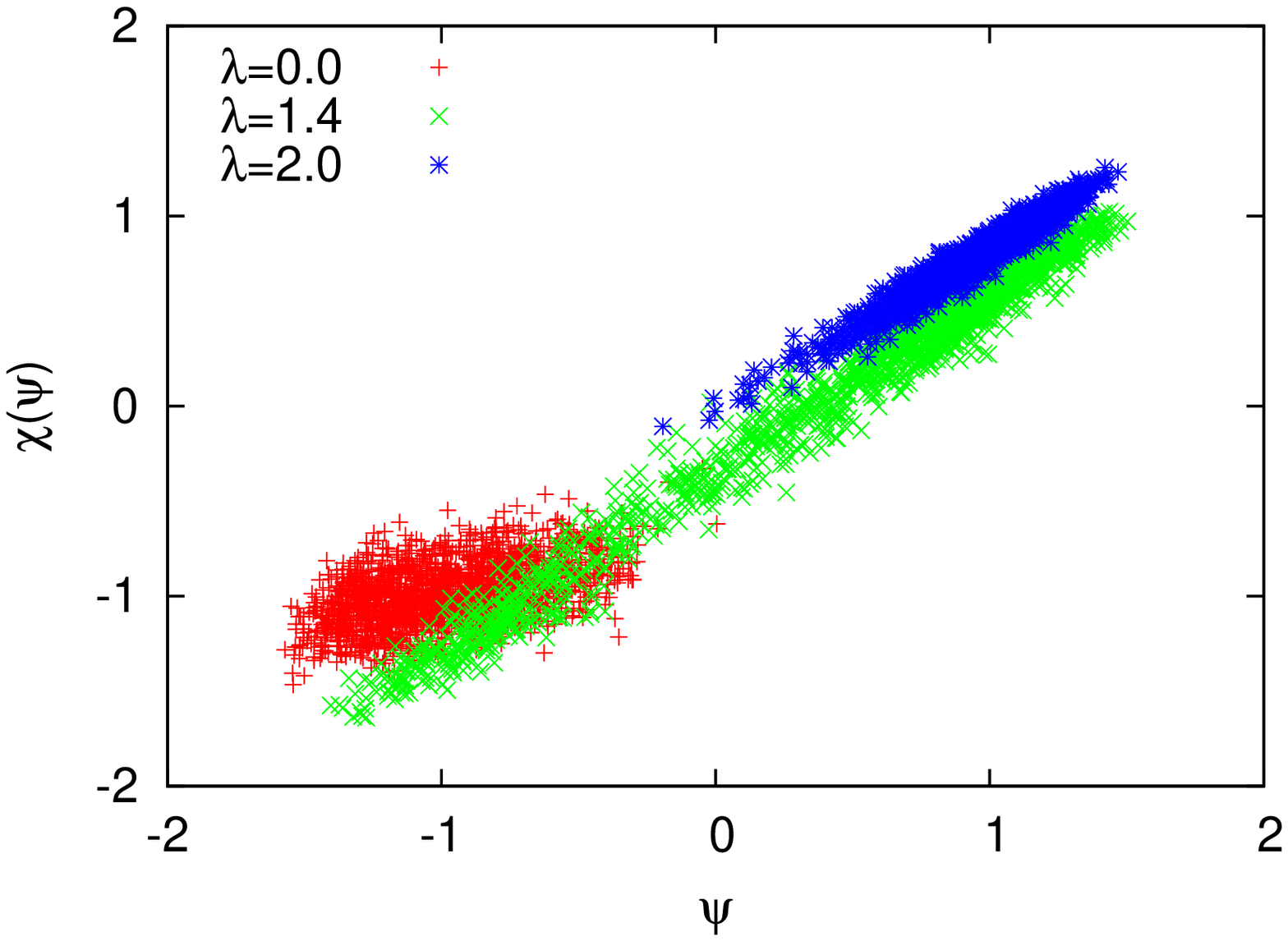}
}
\end{center}
\caption{Effect of spatial correlation strength $\lambda$ on Moran scatterplots of the GL model with additive noise. 
The Moran scatterplots show the effect of the spatial coupling $\lambda$ on the GL model. Panel \subref{A}: scatterplot of the order/disorder/order transition induced by Cai noise ($\delta=-0.5$, $B = 2.6$); panel \subref{B}: Tsallis noise ($q=-1$, $B=2.6$). In all panels $\tau=0.3$. Spatial coupling in both cases increases the slope of scatterplot, and at the same time has a tendency to make the cloud of the scatter plot more compact, lowering the dispersion of short range correlation around the mean value. }
\label{moranscatterplots}
\end{figure}
\section{Moran's scatterplots}
Apart the Moran Index, in order to assess the correlations between a lattice field $\alpha$ and the field $\chi$ constituted by the mean of the values of $\alpha$ in the neighbouring points at each point $p$ of the lattice,  we also employed the scatterplot of the couples $(\alpha_p, \chi(\alpha_p) )$, called Moran scatterplot. This is an useful tool of descriptive spatial statistics. For example, in fig. \ref{fig_3_Tsallis} we show the effects, evaluated by means of the Moran's scatterplots, of the temporal autocorrelation $\tau_c$ on a discretized GL model in a $40\times40$ lattice perturbed by Tsallis noise.